\documentclass[amsmath,amssymb,aps,prb,english,showkeys,twocolumn]{revtex4-2}

\usepackage{graphicx}% Include figure files
\usepackage{dcolumn}% Align table columns on decimal point
\usepackage{bm}% bold math
\usepackage{physics}
\usepackage{hyperref}% add hypertext capabilities
\hypersetup{
    unicode=false,          % non-Latin characters in AcrobatÕs bookmarks
    pdftoolbar=false,        % show AcrobatÕs toolbar?
    pdfmenubar=true,        % show AcrobatÕs menu?
    pdffitwindow=false,     % window fit to page when opened
    pdfstartview={FitH},    % fits the width of the page to the window
    pdfkeywords={MBL} {AGP} {Mass-deformed SYK}, % list of keywords
    pdfnewwindow=true,      % links in new window
    colorlinks=true,       % false: boxed links; true: colored links
    linkcolor=red,          % color of internal links (change box color with linkbordercolor)
    citecolor=blue,        % color of links to bibliography
    filecolor=magenta,      % color of file links
    urlcolor=blue           % color of external links
}
\usepackage{mathtools}
\usepackage{xcolor}
\usepackage[page]{appendix}
\usepackage[normalem]{ulem}

\newcommand{\pcsadd}{Center for Theoretical Physics of Complex Systems, Institute for Basic Science (IBS), Daejeon, Korea, 34126}
\newcommand{\ustadd}{Basic Science Program, Korea University of Science and Technology (UST), Daejeon 34113, Republic of Korea}
\newcommand{\uosakaadd}{Department of Physics, Graduate School of Science, Osaka University, Toyonaka, Osaka 560-0043, Japan}
\newcommand{\sophiaadd}{Physics Division, Sophia University, Chiyoda, Tokyo 102-8554, Japan}

\newcommand{\mh}{\hat{\mathcal{H}}}
\newcommand{\hc}{\hat{\chi}}

\begin{document}

\title{Machine learning wave functions to identify fractal phases
}

\author{Tilen \v{C}ade\v{z}}
    \email{tilencadez@ibs.re.kr}
    \affiliation{\pcsadd}

\author{Barbara Dietz}
    \email{barbara@ibs.re.kr}
    \affiliation{\pcsadd}

\author{Dario Rosa}
    \email{dario_rosa@ibs.re.kr}
    \affiliation{\pcsadd}
    \affiliation{\ustadd}

\author{Alexei Andreanov}
    \email{aalexei@ibs.re.kr}
    \affiliation{\pcsadd}
    \affiliation{\ustadd}

\author{Keith Slevin}
    \email{slevin.keith.sci@osaka-u.ac.jp}
    \affiliation{\uosakaadd}

\author{Tomi Ohtsuki}
    \email{ohtsuki@sophia.ac.jp}
    \affiliation{\sophiaadd}

\date{\today}

\begin{abstract}
    We demonstrate that an image recognition algorithm based on a convolutional neural network provides a powerful procedure to differentiate between ergodic, non-ergodic extended (fractal) and localized phases in various systems: single-particle models, including random-matrix and random-graph models, and many-body quantum systems.
    The network can be successfully trained on a small data set of only \(500\) wave functions (images) per class for a single model.
    The trained network can then be used to classify phases in the other models and is thus very efficient.
    We discuss the strengths and limitations of the approach.
\end{abstract}

%Use showkeys class option if keyword display desired

\maketitle

%\tableofcontents

\section{Introduction}
\label{sec:introduction}

Computational physics has effectively addressed numerous challenges in solid-state physics. 
The recent advances in machine learning techniques~\cite{mitchell2007machine} including deep learning~\cite{lecun2015deep} make it a natural choice for tackling complex problems in physics.
Indeed, since circa 2016~\cite{torlai2016learning,carrasquilla2017machine}, there has been a surge of interest in applying machine learning methods to problems in condensed matter physics~\cite{carleo2019machine}.

One crucial application of machine learning is the extraction of features from data. 
However, electron states in random systems often exhibit intricate features. 
Neural networks obtained through supervised training, which have demonstrated immense potential in image recognition~\cite{lecun2015deep}, are anticipated to be effective for analyzing electron wave behavior in random systems as images~\cite{ohtsuki2020drawing}.

Random free electron systems exhibit the Anderson-type metal-insulator transition, also known as the Anderson transition~\cite{anderson1958absence,kramer1993localization,evers2008anderson}. 
Analyzing the wave functions of random quantum systems can be challenging due to the significant fluctuations in their distribution. 
Nevertheless, a trained convolutional neural network (CNN) has been proven to successfully detect such quantum phase transitions~\cite{ohtsuki2016deep2d,ohtsuki2017deep3d,broecker2017machine,mano2017phase,ohtsuki2020drawing}.
As is well-known in the condensed-matter literature, the Anderson problem, although theoretically very interesting, deals with an idealized situation in which interactions are absent.
Therefore, it is of crucial importance to determine whether Anderson localization is robust when interactions are present and, ultimately, in a genuine quantum many-body setup -- the latter being usually dubbed as \emph{many-body localization} (MBL),~\cite{altshuler1997quasi, basko2006metal, abanin2019colloquium}.
Therefore, it is natural to apply the same CNN approach to detect the \emph{Anderson-like} MBL transition.
Such a task has been addressed from various perspectives and by training the CNN with different kinds of data in the past few years \cite{schindler2017probing, theveniaut2019neural, rao2020machine, kausar2020learning, kotthoff2021distinguishing}.

Some models exhibit localized and extended phases. Anderson's model of localization in 3D and above is a well-known example. 
In contrast with such models, it is generally believed -- although still under intense debate -- that certain models, 
including numerous many-body interacting systems, can exhibit an intermediate phase depending on the 
associated disorder parameter in which eigenstates are neither fully delocalized nor localized.
In the intermediate phase, the states are characterized by non-trivial \emph{fractal} or \emph{multifractal} dimensions.

We have explored the capacity of a CNN to recognize the presence or absence of a fractal phase, both in single particle and in interacting many-body systems, in addition to the localized and extended phases.
To this end, we train the CNN by means of the eigenstates of the generalized Rosenzweig-Porter (gRP) model~\cite{rosenzweig1960repulsion, kravtsov2015random}, a random matrix model, which can be seen as a disordered single-particle system and for which the presence of a fractal phase has been determined analytically~\cite{kravtsov2015random}.
After the CNN has been successfully trained to identify the 
ergodic, fractal and localized phases in the gRP model, we apply the same CNN -- without additional training -- to single particle, random and many-body systems. 
In this way, we assess the capability of the CNN in generalizing the acquired knowledge to new models and situations -- 
a \emph{generalization capability} -- that, 
in the context of MBL, has been studied for the first time in Ref.~\onlinecite{beetar2021neural}.

\begin{figure}[ht]
    \includegraphics[width=1.0\columnwidth]{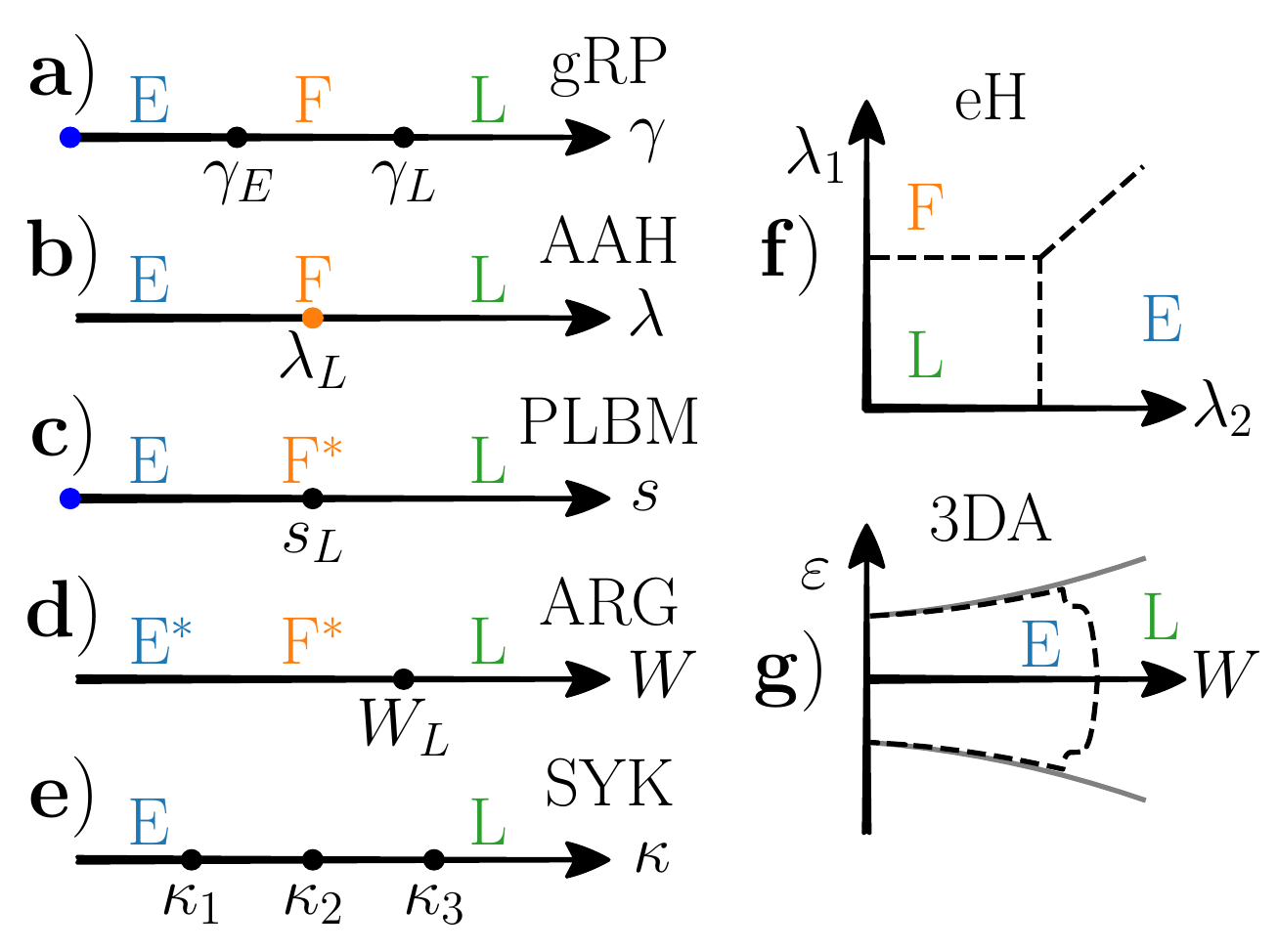}
    \caption{
        Schematic phase diagrams of the considered models. 
        The CNN is trained and tested on the gRP model. 
        The corresponding phase diagram is shown in a). 
        The same network is then applied to 
        b) the Aubry-Andr\'e-Harper model (AAH), 
        c) the power-law banded matrices (PLBM), 
        d) the Anderson model on random graphs (ARG), 
        e) the mass deformed Sachdev-Ye-Kitaev model (SYK), 
        f) the extended Harper's model (eH) and 
        g) the 3D Anderson model (3DA). 
        In the left and right panels the dots and dashed lines, respectively, mark known phase transitions (see main text). 
        Blue dots mark the parameter value where the model Hamiltonian is a member of the GOE. 
        The orange dot in b) signifies that (multi-) fractal states are present in the AAH only at the transition point. 
        The letters E,F,L denote the ergodic, fractal, localized phases, respectively, which are the output classes of the CNN. 
        The asterisk in the phase diagrams of the PLBM model and the Anderson model on random graphs signifies that the corresponding phases are observed in the finite-size systems that are considered.
    }
    \label{fig:schematic_phase_diagrams}
\end{figure}

The paper is organized as follows. 
The methods, namely the CNN and the exact diagonalization are introduced in Sec.~\ref{sec:methods}. 
The various models of interest are discussed in Sec.~\ref{sec:models} and their phase diagrams are schematically shown in Fig.~\ref{fig:schematic_phase_diagrams}. 
The generalized Rosenzweig-Porter model, introduced in Sec.~\ref{subsec:gRP}, is used to train and test the CNN, as given in Sec.~\ref{subsec:CNN_training}. 
That network is then applied to other models in Sec.~\ref{subsec:application}. 
Finally, the results are discussed and the conclusions are given in Sec.~\ref{sec:conclusions}.

\section{Methods}
\label{sec:methods}

Convolutional neural networks are networks which take a certain input in the form of a single or multiple arrays, process it and produce an output, based on the task to be fulfilled. 
In the case of image recognition the CNN takes an image as an input and as an output classifies the content of the image. 
A famous example is the handwritten digit recognition~\cite{lecun1989handwritten}, where the task is to correctly recognize handwritten digits. 
Here we use the CNN to recognize different phases of matter~\cite{ohtsuki2016deep2d,carrasquilla2017machine,broecker2017machine}, specifically the ergodic extended phase, the non-ergodic extended (fractal) phase and a localized phase in various systems. 
Typical CNNs are composed of several main parts: convolutional layers, pooling layers and standard dense layers. 
The first two of these offer an improvement compared to the simpler artificial neural networks, consisting of only dense layers, while also reducing the number of parameters to be optimized. 
We refer the interested reader to Ref.~\onlinecite{ohtsuki2020drawing} for further details on supervised learning of disordered quantum systems.

To obtain the eigenfunctions of various models we use the exact diagonalization by solving the equation \(H \ket{\psi_{\mu}} = \varepsilon_{\mu} \ket{\psi_{\mu}}\), which is written in the computational basis \( \ket{\psi_{\mu}} = \sum_i \psi_{\mu}(i) \ket{i}\) and \( \psi_{\mu}(i)\) are the corresponding coefficients of the eigenfunction. 
The input data given to the CNN are the squares of the absolute value of the eigenstate coefficients  \(\abs{\psi_{\mu}(i)}^2\) (the probabilities of site occupations). 
For comparison we also calculate the \emph{inverse participation ratio} (IPR) \({\mathcal{I}}\), given by \({\mathcal{I}} = \langle \sum_i \abs{\psi_{\mu}(i)}^{4} \rangle\), where the average can be either over the eigenstates \(\mu\) in a chosen energy window or over different disorder realizations or both.

\section{Models and results}
\label{sec:models}

We consider a set of models that exhibit a transition from extended to localized phases.
In some of these models the transition takes place via an intermediate fractal phase.
In other models, such as Anderson's model of localization, a fractal regime (as opposed to a phase) can be observed as a consequence of a finite-size effect.

\subsection{Generalized Rosenzweig-Porter model (\textbf{gRP})}
\label{subsec:gRP}

To train and test the CNN we use the eigenstates of the gRP model~\cite{rosenzweig1960repulsion, kravtsov2015random}, which comprises Hermitian random matrices whose elements are Gaussian distributed with zero mean. 
The variances of the diagonal and off-diagonal elements, denoted by \(\sigma_d^2\) and \(\sigma_{off}^2\), respectively, are defined as
\begin{align}
    \label{eq:gRP_variances}
    \sigma_d^2 = \langle H_{nn}^2 \rangle = \frac{1}{2 N}, \quad \sigma_{off}^2 = \langle H_{nm}^2 \rangle = \frac{1}{4 \, N^{\gamma + 1}}.
\end{align}
Here, the parameter \(\gamma\) determines the strength of the off-diagonal matrix elements compared to that of the diagonal ones.
In this work we consider real matrices implying that for \(\gamma = 0\) they are members of the Gaussian orthogonal ensemble (GOE). 
In Ref.~\onlinecite{kravtsov2015random} the phase diagram, which is schematically shown in panel a) of Fig.~\ref{fig:schematic_phase_diagrams}, was established. 
It was shown that the states around the band center exhibit three distinct phases, namely an ergodic phase for \(\gamma < 1\), an extended non-ergodic phase for \(1 < \gamma < 2\) and a localized phase for \(\gamma > 2\). 
At \(\gamma_E = 1\) and \(\gamma_L = 2\) they undergo  continuous ergodic and Anderson transitions, respectively. 
The characteristics of the recently discovered extended non-ergodic phase is the occurrence of fractal eigenstates whose fractal dimension equals \(2-\gamma\). 
The phase diagram was confirmed and the properties of the model were further studied recently~\cite{landon2019fixed, facoetti2016from, truong2016eigenvectors, monthus2017multifractality, vonsoosten2019non, bogomolny2018eigenfunction, pino2019from, detomasi2019survival, berkovits2020super, skvortsov2022sensitivity}. 
The ergodic and localized phases can also be determined using simple criteria~\cite{bogomolny2018power, khaymovich2020fragile}, as shown in Appendix~\ref{app:criteria}. 
Here we use the three distinct phases of the  gRP model as the output classes of the CNN. 

\subsection{CNN training and testing on the gRP model}
\label{subsec:CNN_training}

For training the CNN we use the eigenstates obtained from diagonalizing the gRP model. 
We use \(N \times N\) matrices with \(N = 2048\) and provide the absolute-value square of the eigenstate coefficients, that is, the occupation probabilities of the sites in the computational basis, to the input layer. 
For each random-matrix realization we extract a single eigenstate corresponding to the eigenenergy closest to the band center, which is at energy \(0\), and we use \(500\) such eigenstates for each of the three phases as input training data set.
During the training \(90 \%\) of the input data is used as a training set and the remaining \(10 \%\) as the validation set.
We observed that the performance improved when applying for all the eigenfunctions a cyclic permutation to the component indices such that the maximum occupation is at the center of the computational basis. 
Accordingly, assuming periodic boundaries, we applied this procedure to all the data considered.
The output layer classifies the ergodic, fractal, and localized phases in terms of probabilities for each phase. 
The objective of this work is to first use the CNN to classify these three phases in the trained model itself and then to apply the same network to various systems: single-particle models, including random matrix models, random graphs models, and many-body quantum systems.
In this way, we will test the ability of CNNs to serve as \emph{diagnostic} tool, \emph{i.e.} as a tool to uncover the presence of interesting phase diagrams in new and unknown models.

The network architecture consists of two convolutional layers, each followed by a pooling layer where we utilize a max pooling strategy. 
We flatten the data and apply a dense layer after the second pooling layer, followed by a rectified linear unit (ReLU) activation function. 
Finally a second dense layer is applied followed by a softmax activation~\cite{ohtsuki2020drawing}. 
Further details and the network hyperparameters are provided in Appendix~\ref{app:CNN}. 
We use the categorical crossentropy between the output probabilities as a loss function during the training for each of the three phases and the corresponding labels. 

\begin{figure}
    \includegraphics[width=0.75\columnwidth]{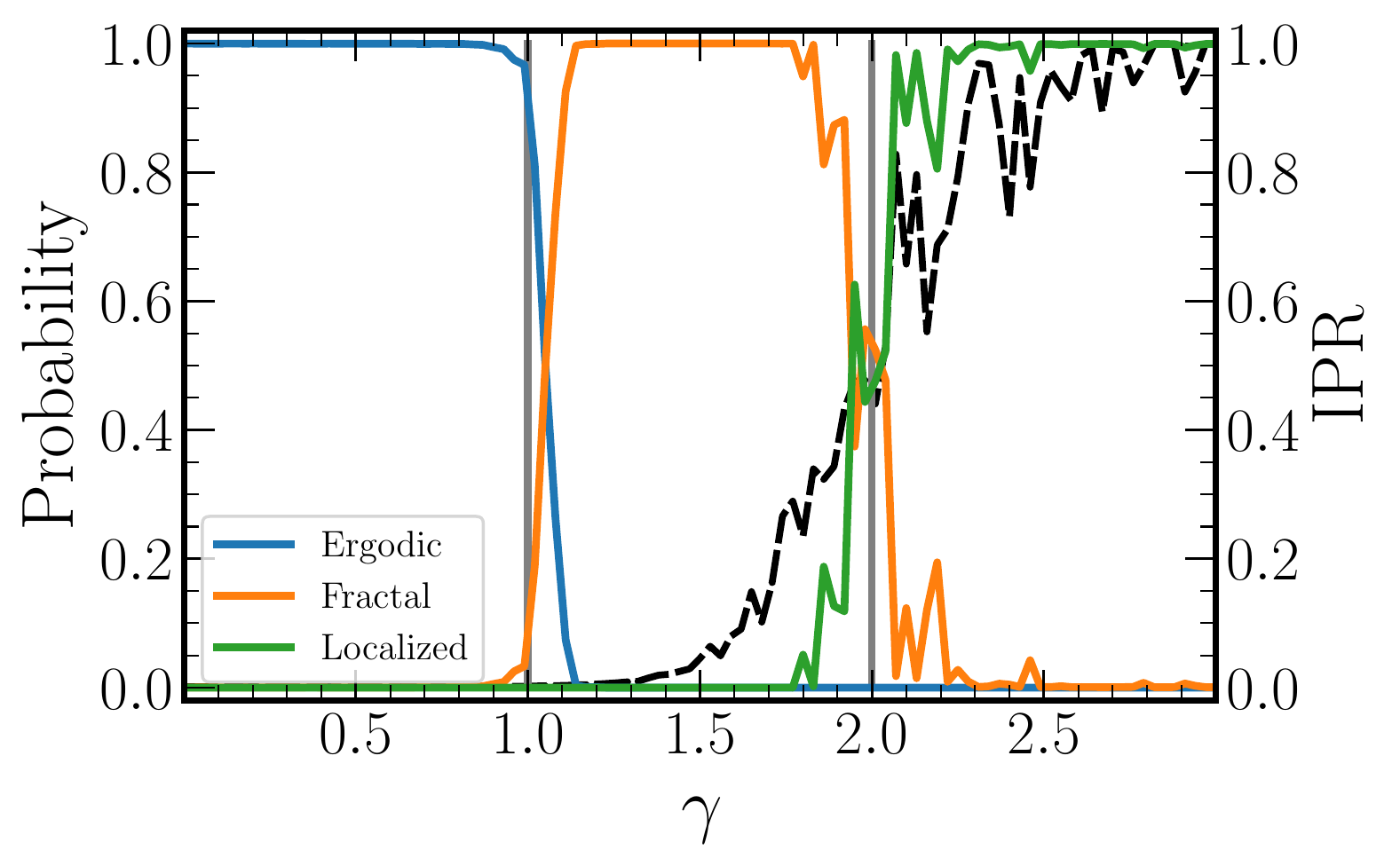}
    \caption{
        Testing the trained CNN on the gRP model. 
        The probabilities of each of the phases as well as the average IPR (black dashed line) are shown. 
        The gray vertical lines indicate the analytical values \(\gamma_E\) and \(\gamma_L\) for the ergodic and Anderson transitions.
    }
    \label{fig:gRP_testing}
\end{figure}

After the training, we test the CNN on a new set of data generated as follows.
We consider a sequence of values of \(\gamma\in [0, 3]\) in which \(\gamma\) is increased in steps of \(0.03\).
For each \(\gamma\) we generate \(5\) different random matrix realizations.
For each realization we input to the CNN the eigenstate with energy closest to the band center and average the probabilities output by the CNN over the \(5\) realizations.
The resulting probabilities for the three phases are shown in Fig.~\ref{fig:gRP_testing}.
The CNN successfully recognizes each of the phases with probability close to \(1\). 
For both the phase transitions, the precision of the determination of the critical value of \(\gamma\) is about 10\%.
For comparison, we also plot the average IPR. 
We see that, for a given system size and pool of eigenstates, the different phases are more clearly discernible with the CNN.

\subsection{Generalization capability - application to other models}
\label{subsec:application}

Having demonstrated that a CNN can successfully classify the phases of the gRP model, we next demonstrate a \emph{generalization capability}:
A CNN trained on one model -- the gRP model -- can classify phases of other models, \emph{without any retraining}. 

\subsubsection{Aubry-Andr\'{e}-Harper model \normalfont{(\textbf{AAH})}}

The Aubry-Andr\'{e}-Harper model~\cite{aubry1980analyticity} is a 1D single-particle model of a particle hopping along a tight binding chain with homogeneous hopping (set to \(1\)) and a quasi-periodic onsite potential, with the Hamiltonian given by
\begin{align}
    \label{eq:AAH}
    H \psi_n = V_n \psi_n + \psi_{n+1} + \psi_{n-1}.
\end{align}
The potential at site \(n\) is \(V_n = \lambda \cos(2 \pi \alpha n + \varphi)\), with a modulation parameter \(\alpha\) chosen as an irrational number. 
We chose for its value the inverse golden mean \(\alpha = (\sqrt{5}-1)/2\). 
The angle \(\varphi \in [ 0, 2 \pi )\) is a simple phase shift of the modulation. 
We use it to generate multiple disorder realizations in finite-size systems. 
The model has two phases, metallic with extended eigenstates for \(\lambda < 2\) and insulating with localized eigenstates for \(\lambda > 2\). 
At the Anderson transition (\(\lambda_L = 2\)) the eigenstates are multi-fractal. 
Furthermore, the two phases are connected via a duality transformation as can be seen by looking at the Fourier transform of the Hamiltonian in Eq.~\eqref{eq:AAH}.
All the eigenstates have the same characteristic length, given by the localization length \(\xi = 1/\log(\lambda/2)\) where \(\lambda > 2\)~\cite{aubry1980analyticity}.
The phase diagram is shown in panel b) of Fig.~\ref{fig:schematic_phase_diagrams}. 

\begin{figure}
    \includegraphics[width=0.75\columnwidth]{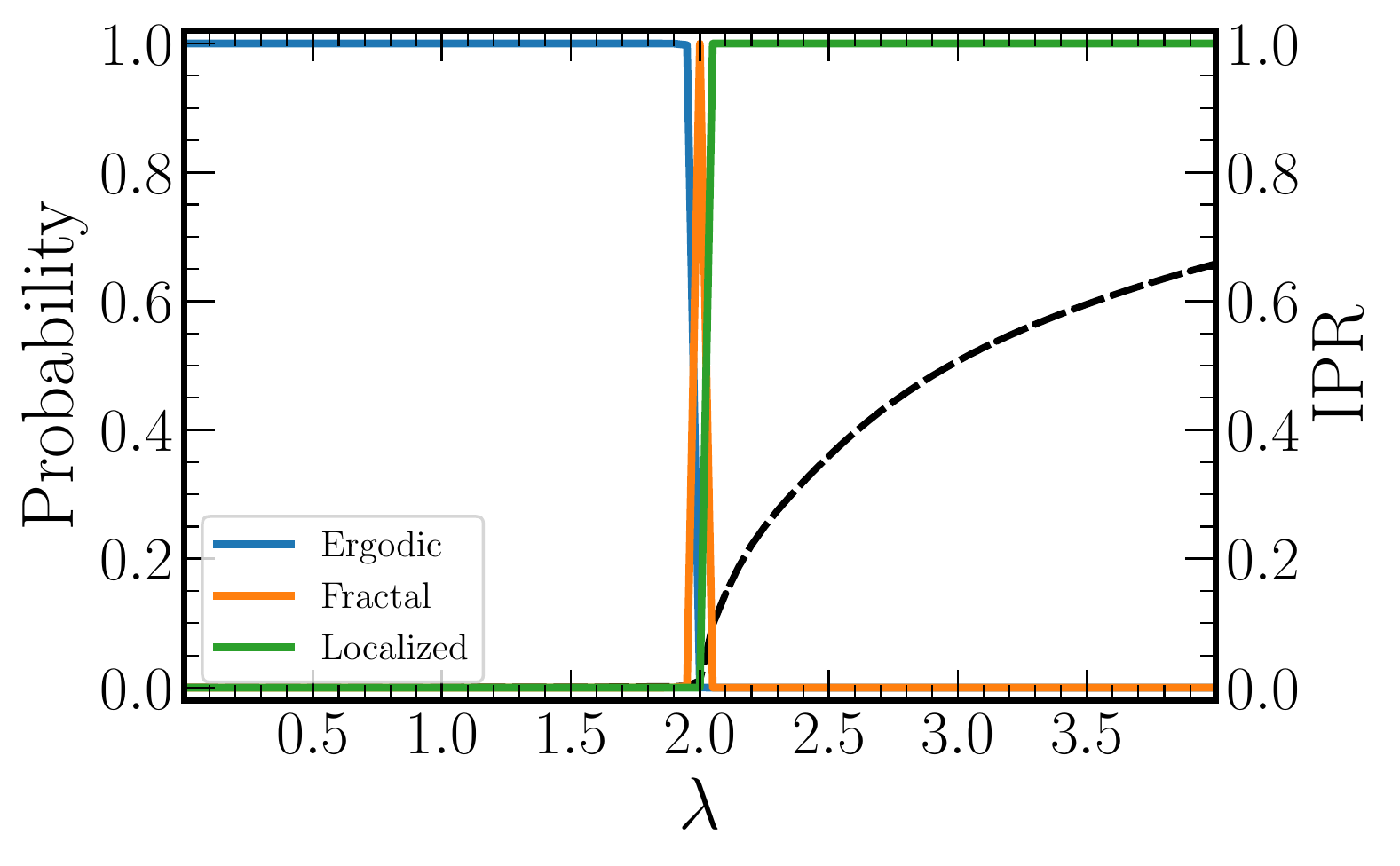}
    \caption{
        Testing the trained CNN on the AAH model. 
        The probabilities of each of the phases are shown as well as the average IPR (dashed line).
        The Anderson transition takes place at \(\lambda_L=2\).
    }
    \label{fig:AAH_testing}
\end{figure}

\begin{figure}
    \includegraphics[width=0.75\columnwidth]{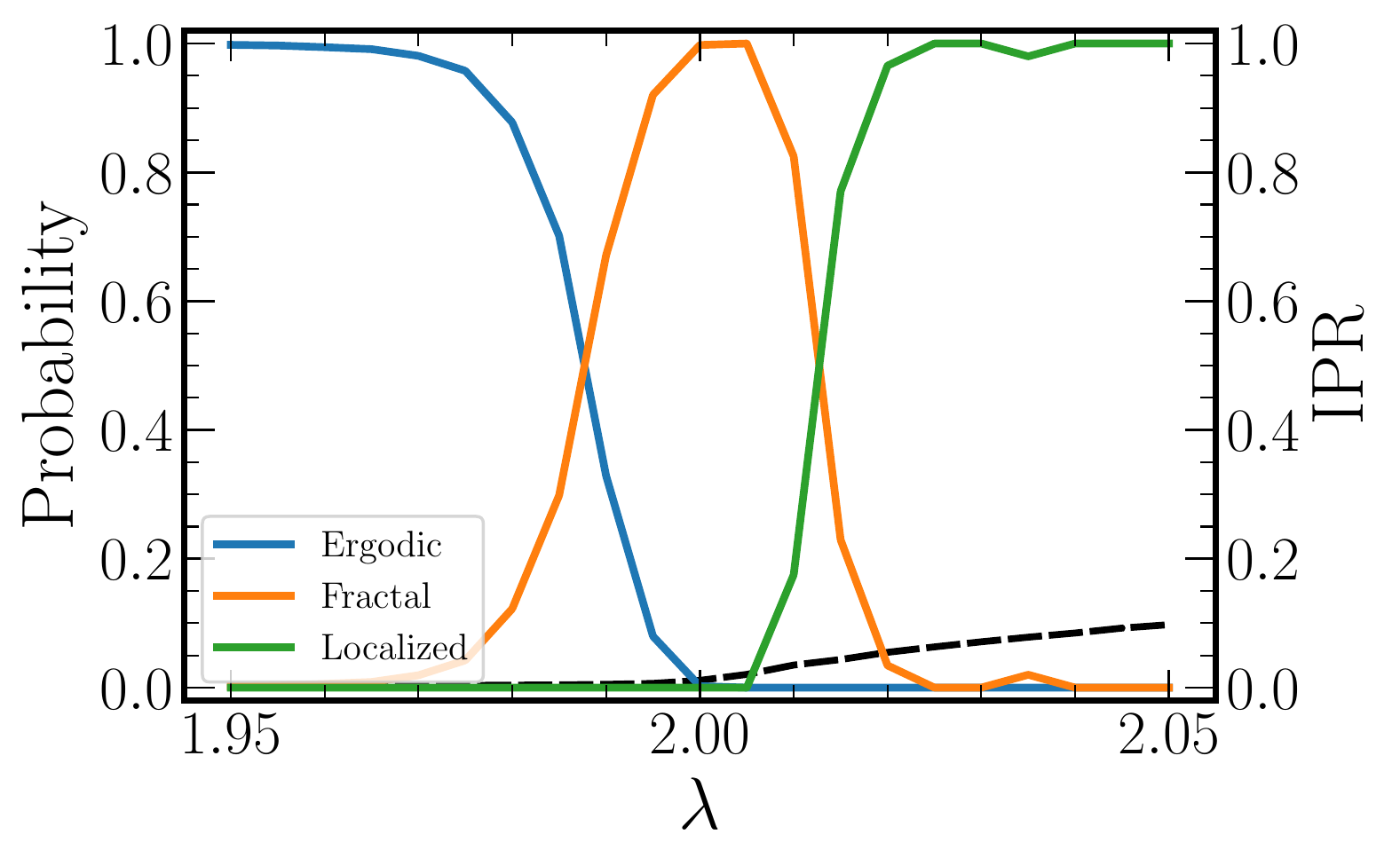}
    \caption{
        Zoom into the transition region of the AAH model.
        The probabilities of each of the phases are shown as well as the average IPR (dashed line).
    }
    \label{fig:AAH_zoom}
\end{figure}

We test the CNN, previously trained on the gRP model, as follows.
We consider a sequence of values of \(\lambda\in [0, 4]\)\ in which \(\lambda\) is increased in steps of \(0.05\).
For each \(\lambda\) we generate \(5\) quasi-disorder realizations.
For each realization we input to the CNN the eigenstate with energy closest to the energy \(0\) and average the probabilities output by the CNN over the \(5\) realizations.
We use the same system size \(N = 2048\) as for the gRP model and impose open boundary conditions.
As can be seen in Fig.~\ref{fig:AAH_testing}, the CNN recognizes the two distinct phases and also correctly identifies the transition point itself as being fractal, all with probabilities close to \(1\). 
The transition region is sharper than in the gRP model. 
To get a clearer picture of its structure, we reduced the \(\lambda\) step to \(0.005\) and increased the number of
realizations to \(50\). 
The results are plotted in Fig.~\ref{fig:AAH_zoom}.
The curves for the localized and fractal phases cross at approximately \(\lambda=2.012\).
For this value of \(\lambda\), the localization length is \(\xi \approx 167\). 
This is approximately one order of magnitude smaller than the system size \(N\).

\subsubsection{Extended Harper's model \normalfont{(\textbf{eH})}}

For the extended Harper's model (eH) the phase diagram was recently established analytically~\cite{avila2017spectral} and is schematically shown in panel f) of Fig.~\ref{fig:schematic_phase_diagrams}. 
It is a one dimensional nearest-neighbor hopping model. 
The Hamiltonian is given by
\begin{align}
    \label{eq:ext_Harper}
    H \psi_n & = V_n \psi_n + t_n \psi_{n+1} + t_{n-1}^* \psi_{n-1}, \\
    V_n & = 2 \, \cos(2 \pi \alpha n + \varphi) \notag \\
    t_n & = \lambda_1 \, e^{-2 i \pi \alpha (n + 1/2) -i \varphi} + \lambda_2 + \lambda_3 \, e^{2 i \pi \alpha (n + 1/2) + i \varphi} \notag
\end{align}
with the modulated onsite potential \(V_n\) and modulated nearest neighbor hoppings \(t_n\).
As in the AAH model we choose \(\alpha = (\sqrt{5}-1)/2\).
In studies of the eH model, disorder realizations are often generated by taking \(\varphi\) to be
randomly distributed on \([ 0, 2 \pi )\).
However, below we consider only a single realization and accordingly set \(\varphi=0\).
The three parameters \(\lambda_1, \lambda_2, \lambda_3\) determine the phase diagram of the model. 
Without loss of generality we restrict to the cases \( \lambda_2 \ge 0, \lambda_1 + \lambda_3 \ge 0\) and \(\lambda_i > 0\) for at least one of \( i = 1,2,3\). We consider the symmetric case \( \lambda_1 \equiv \lambda_3\) where \(t_n = 2 \lambda_1 \, \cos[2 \pi \alpha (n + 1/2)] + \lambda_2\) and the Hamiltonian in Eq.~\eqref{eq:ext_Harper} is purely real.

It was shown in Ref.~\onlinecite{avila2017spectral} that in the symmetric case of a purely real Hamiltonian the spectrum of the eH model belongs to one of the three distinct cases:
\begin{itemize}
    \item Region I: localized eigenfunctions / pure point spectrum: \\
    \(0 \le \lambda_1 + \lambda_3 \le 1,\, 0 < \lambda_2 \le 1\)
    \item Region II: extended eigenfunctions / purely absolutely continuous spectrum: \\
    \(0 \le \lambda_1 + \lambda_3 \le \lambda_2,\, \lambda_2 \ge 1\)
    \item Region III: fractal eigenfunctions / purely singular continuous spectrum: \\
    \(\max{\{1, \lambda_2\}} \le \lambda_1 + \lambda_3, \, \lambda_2 > 0\)
\end{itemize}
The transition lines between the three regions exhibit fractal eigenstates. 
In the non-symmetric case, that is, for \(\lambda_1 \ne \lambda_3\), regions I and II persist, whereas the eigenstates of region III become extended.

\begin{figure}
    \includegraphics[width=0.7\columnwidth]{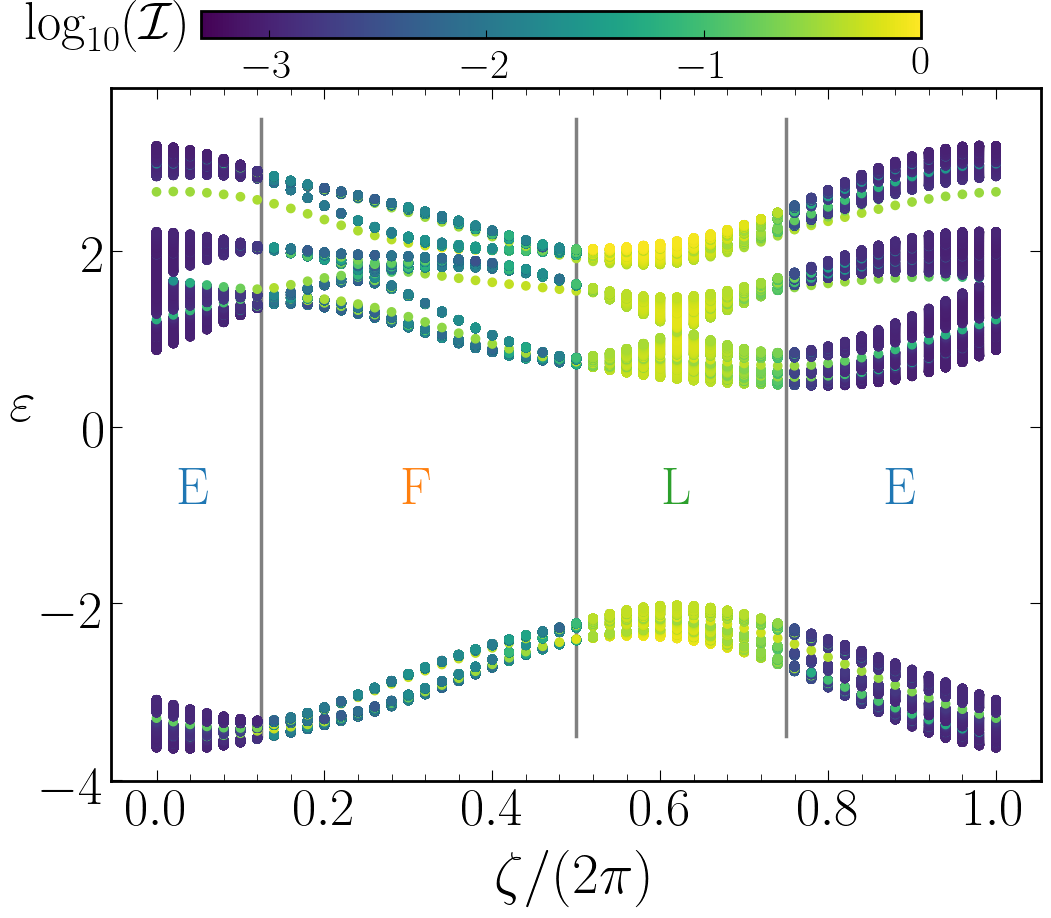}
    \includegraphics[width=0.7\columnwidth]{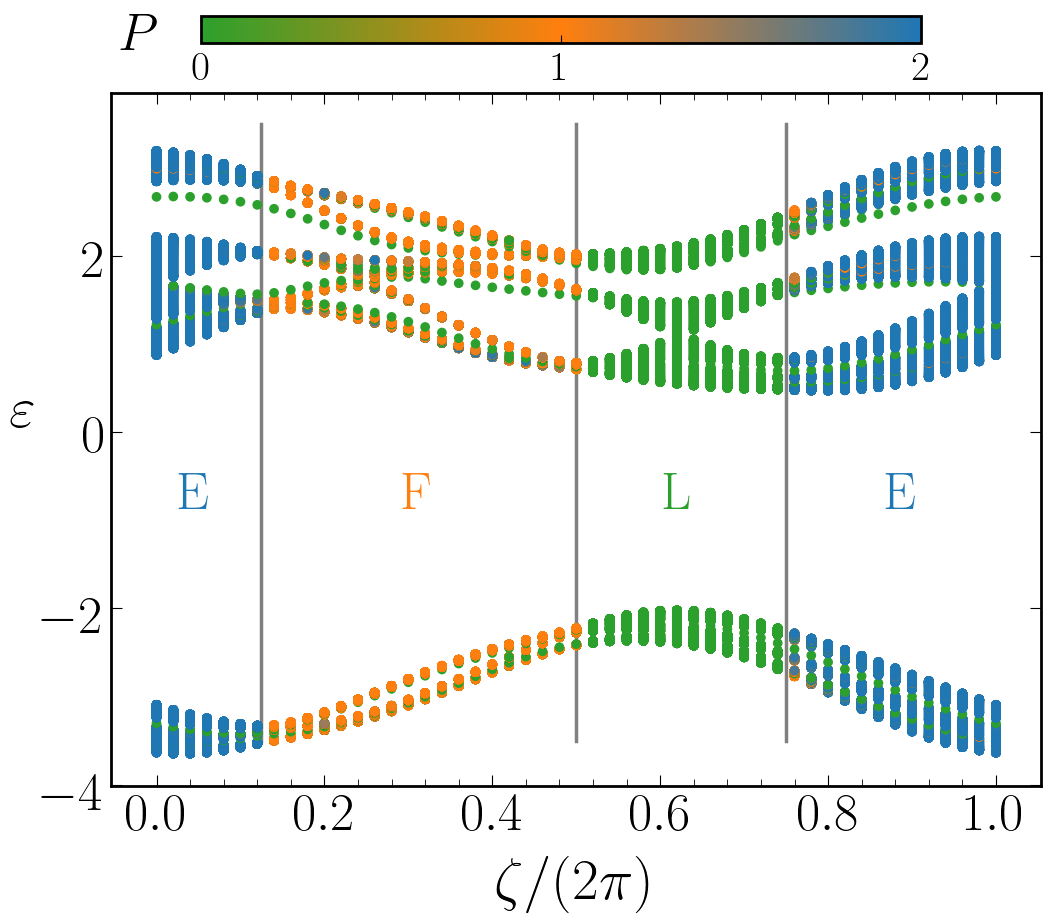}
    \caption{
        Testing the trained CNN on the eH model. 
        Upper panel: Logarithm of the IPR while traversing a loop in the parameter space for the full energy spectrum \(\varepsilon\). 
        Lower panel: A simple function of the probabilities associated with the different phases obtained from the CNN, \(P = 0\cdot P_L + 1\cdot P_F + 2\cdot P_E\), where \(P_L, P_F, P_E\) are the probabilities for localized (L), fractal (F) and ergodic (E) phases, respectively~\cite{ohtsuki2017deep3d}. 
        Gray vertical lines mark the transition values. 
        Note the presence of localized states in the ergodic and fractal phases. 
        They originate from the open boundary conditions.
    }
    \label{fig:extended_Harper}
\end{figure}

To reduce the number of parameters, we consider a closed loop in the two parameter space \(\lambda_1, \lambda_2\), which transverses all the three distinct phases of the model. 
We parameterized the loop by an angle \(\zeta\) so that
\begin{align}
    \label{eq:extended_Harper_loop}
    \lambda_1 & = 0.5 + r_0 \sin(\zeta), \nonumber \\ 
    \lambda_2 & = 1.0 + 2 r_0 \cos(\zeta).
\end{align}
We set \(r_0 = 1/4\) and increased \(\zeta\) from \(0\) to \(2\pi\) in steps of \(0.02 \pi\).
For each point on the loop, we test the CNN on \emph{the full spectrum} of eigenfunctions. 
We use open boundary conditions, for which localized edge states appear in the band gaps.
The results are presented in Fig.~\ref{fig:extended_Harper}. 
The CNN recognizes the ergodic and localized phases with probabilities close to \(1\) for the majority of the states and correctly identifies about \(85 \%\) of the fractal states. 
The transition regions are sharp. 
The localized edge states are also successfully identified by the CNN.

Recently a related model, called the generalized AAH model, was studied~\cite{he2022persistent} using a topological machine learning technique of persistent homology. 
The authors successfully distinguished the localized, extended, and critical phases within the model.

\subsubsection{Three-dimensional Anderson model \normalfont{(\textbf{3DA})}}

A paradigmatic model of metal-insulator transitions is Anderson's model of localization~\cite{anderson1958absence,kramer1993localization,evers2008anderson}. 
It describes a single particle hopping on a tight binding lattice.
We consider a three-dimensional (3D) cubic lattice, with homogeneous hopping (set to unity) and a random onsite potential disorder. 
The onsite potentials are independently and uniformly distributed on \([-W/2, W/2]\). 
The parameter \(W\) defines the disorder strength. 
The model has been extensively studied numerically~\cite{pichard1981finite,mackinnon1981one,mackinnon1983the,
slevin1999corrections,rodriguez2010critical,rodriguez2011multifractal,slevin2014critical,slevin2018critical,suntajs2021spectral}. 
Its phase diagram is shown schematically in panel g) of Fig.~\ref{fig:schematic_phase_diagrams}.
At the phase boundary, an Anderson transition, where eigenstates are multi-fractal, separates localized and extended eigenstates.
In Refs.~\onlinecite{ohtsuki2017deep3d, mano2017phase} the 3DA was recently studied using the CNN, and the authors demonstrated that the CNN can efficiently recognize the metallic and Anderson localized phases.

We simulate a \(16 \times 16 \times 8\) lattice and impose periodic boundary conditions.
We increase \(W\) through the range \([0,35]\) in steps of \(0.5\).
For each \(W\), we generate a single realization and compute the entire spectrum by exact diagonalization.
While the lattice has \(N=2048\) states, which matches the input layer of the CNN, we need to flatten the data.
In doing so, some spatial information is lost.

\begin{figure}
    \includegraphics[width=0.73\columnwidth]{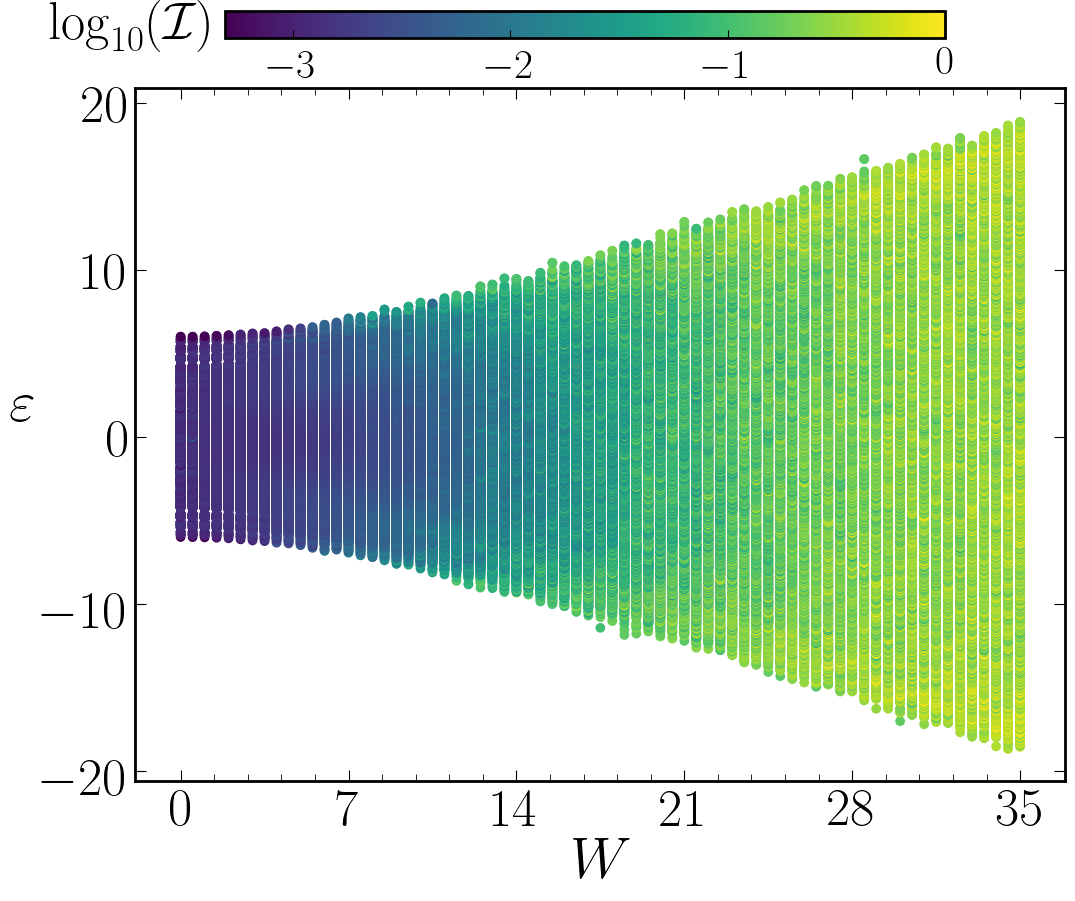}
    \includegraphics[width=0.73\columnwidth]{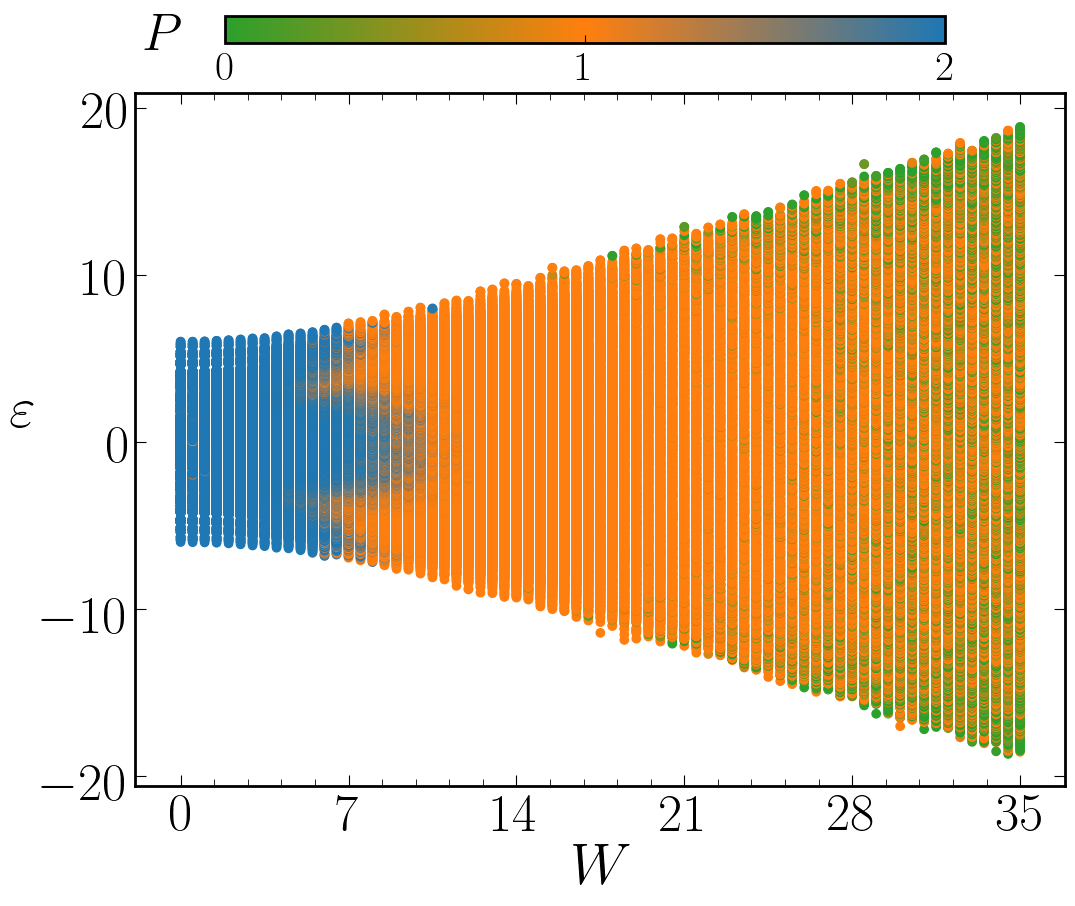}
    \caption{
        Testing the trained CNN on the 3DA model. 
        Upper panel: The logarithm of the IPR for different disorder strengths \(W\) and for the full energy spectrum \(\varepsilon\). 
        Lower panel: Probability \(P\) for the different phases obtained using CNN. 
        Here, \(P\) is defined in the caption of Fig. \ref{fig:extended_Harper}. 
        Note the very strong finite-size effects (orange colored points in the lower panel) attributed to the small system size of \(16\times 16\times 8\) yielding \(N = 2048\)).
    }
    \label{fig:3D_Anderson}
\end{figure}

In Fig.~\ref{fig:3D_Anderson} we show the results. 
The CNN correctly recognizes the ergodic phase for \(W < 7\), however it incorrectly classifies most of the states for \(7 < W < 30\) as being fractal and only at the stronger disorder \(W > 30\) the eigenstates are gradually identified as localized.

One possible explanation for this failure is that it is a finite size effect.
At the band center \(\epsilon=0\), the Anderson transition occurs at a critical disorder \(W_c = 16.54 \pm 0.01\)~\cite{slevin2014critical}.
Estimates for the correlation, respectively, localization length for the band center have been tabulated in Ref.~\onlinecite{mackinnon1983the}.
The correlation length is approximately \(1\) lattice spacing for \(W=10\) and the localization length is approximately \(2\) lattice spacings for \(W=30\).
Given the results above for the gRP model and the dimensions of the systems simulated here, an explanation in terms of a finite size effect is plausible.
Another possibility is that the failure is due to the loss of spatial information that results from the flattening of the data that is dictated by the gRP model, 
which was used to train the CNN, and for which there is no concept of a spatial lattice.

\subsubsection{Power-law random banded matrices \normalfont{(\textbf{PLBM})}}

The power-law random banded matrices~\cite{mirlin1996transition} that we consider are real symmetric matrices with Gaussian distributed elements with zero mean and variances
\begin{align}
    \label{eq:PLBM_variances}
    \sigma_d^2 = \langle H_{nn}^2 \rangle = \frac{1}{2 N}, \quad \sigma_{off}^2 = \langle H_{nm}^2 \rangle = \frac{1}{4 \, N} a^2(\abs{n-m}).
\end{align}
Here, \(N\) is the matrix dimension and \(a(|n-m|)\) is a function of the distance \(r = |n-m|\) from the diagonal, which at large distances decreases according to a power law, \(a(r) \sim r^{-s}\) for \(r \gg 1\) with \(s \ge 0\). 
We adopt the function \(a(r)\) introduced in the original work on power-law random banded matrices~\cite{mirlin1996transition},
\begin{align}
    \label{eq:PLBM_a}
    a(r) = 
    \begin{cases}
        1 & r \le b, \\
        (\bar{r}/b)^{-s} & r > b,
    \end{cases}
\end{align}
where we define \(\bar{r} = {\mathrm{min}}(r, N-r)\) to reduce boundary effects and \(b\) is an additional parameter. 
It was shown analytically~\cite{mirlin1996transition} that for \(b \gg 1\) the model exhibits a phase transition from extended to localized eigenstates as a function of \(s\) at \( s=1 \). 
However the case \(b=1\) shows an anomalously large critical region around the transition point \( s_L=1 \)~\cite{cuevas2001anomalously} where multi-fractal eigenstates persist up to extremely large system sizes. 
We focus on that case.

\begin{figure}
    \includegraphics[width=0.75\columnwidth]{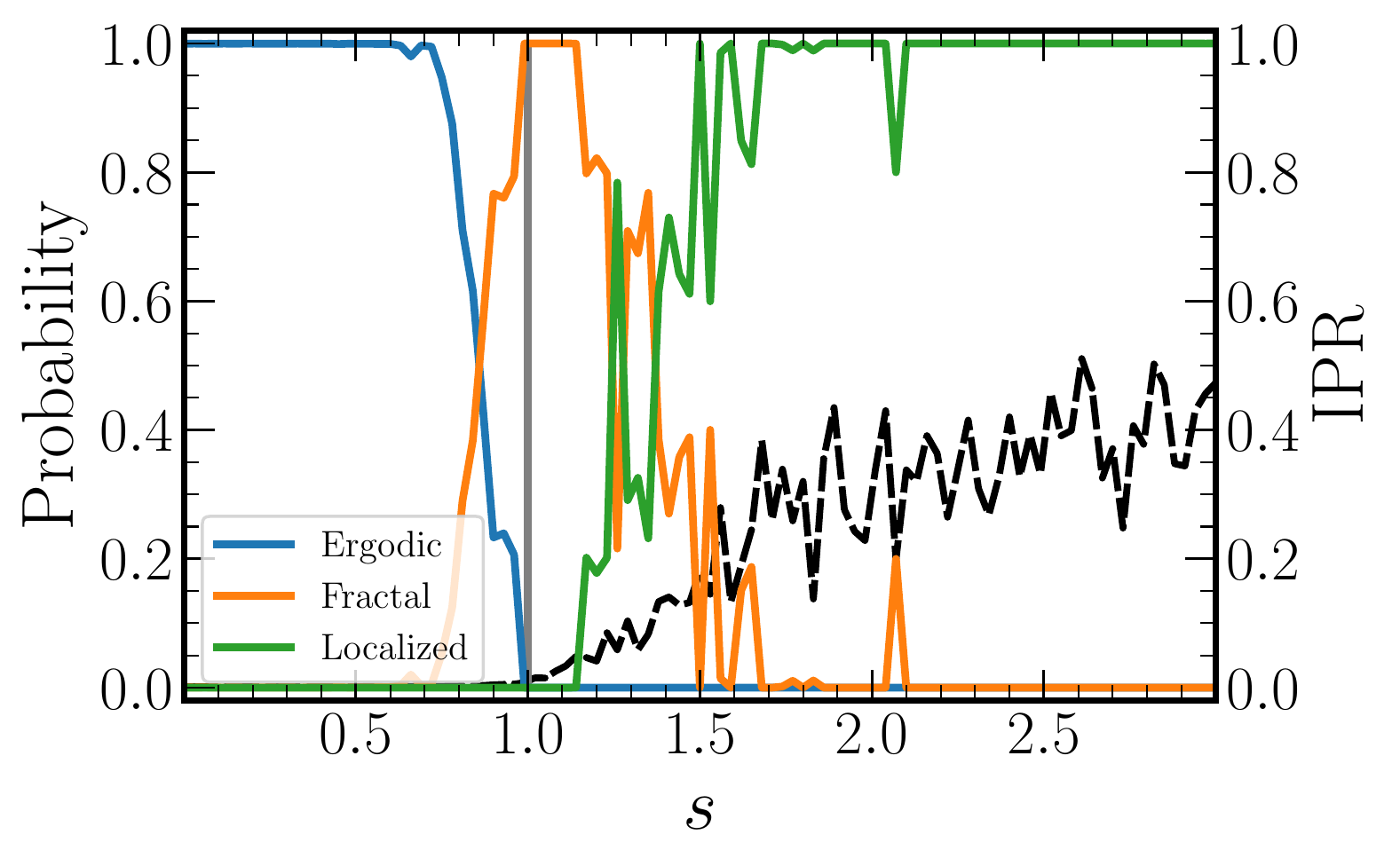}
    \caption{
        Testing the trained CNN on the PLBM model.
        The probabilities of each of the phases are shown as well as the average IPR (black dashed line). 
        The gray vertical line marks the analytical value \(s_L\) for the transition from extended to localized states.
    }
    \label{fig:PLBM_testing}
\end{figure}

We test the CNN as follows. 
We increase \(s\) through the range \([0,3]\) in steps \(0.03\).
For each \(s\) we generate \(5\) realizations, extract a single eigenstate closest to the band center and input them in turn to the CNN.
We then average the probabilities output by the CNN over the \(5\) realizations.
The probabilities for each of the phases and the average IPR are shown in Fig.~\ref{fig:PLBM_testing}. 
In good agreement with other numerical results~\cite{cuevas2001anomalously} the CNN recognizes the ergodic and localized phases, with the intermediate fractal regime.

\subsubsection{Anderson model on random graphs \normalfont{(\textbf{ARG})}}

\begin{figure}
    \includegraphics[width=0.45\columnwidth]{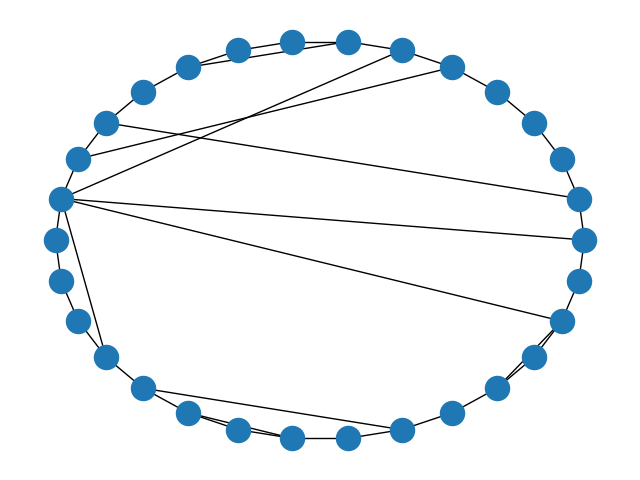}
    \includegraphics[width=0.45\columnwidth]{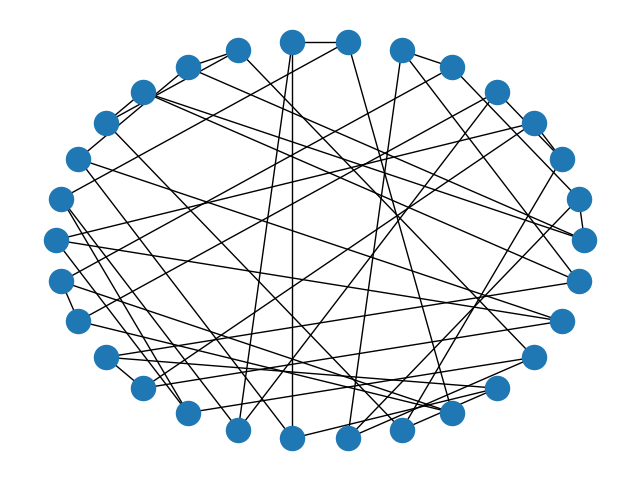}
    \caption{
        Examples of the small-world network with \(p = 0.25\) (left) and an RRG (right), both have \(30\) nodes.
    }
    \label{fig:random_graphs}
\end{figure}

The Anderson model on random graphs has been extensively studied recently~\cite{deluca2014anderson, altshuler2016nonergodic, kravtsov2018non, tikhonov2016anderson, tikhonov2019critical, biroli2022critical,  pino2020scaling, garcia-mata2017scaling, garcia-mata2020two, garcia-mata2022critical}.
Here we consider two types of graphs, the first is a variant of the celebrated small-world networks~\cite{milgram1967small,watts1998collective,newman2000mean} and the second is the random regular graph (RRG). 
The corresponding Hamiltonian describing a single particle on a tight binding lattice with onsite potential disorder, where the lattice is defined by the choice of the graph, can be written as
\begin{align}
    \label{eq:AM_RRG}
    H \psi_n = \varepsilon_n \psi_n + \sum_{m} A_{nm} \psi_{m}.
\end{align}
Here \(\varepsilon_n\) is the onsite potential and \(A\) is the adjacency matrix of the random graph with hopping set to \(1\).
For the small-world network~\cite{newman2000mean} we use the nearest neighbour hopping and add additional long range connections among two random sites with probability \(p\), 
so that the average node degree is \(2+2 p\).
Note that the case \(p=0\) corresponds to the 1D Anderson model.
For the RRG the node degree is a fixed number \(v\), which we choose to be \(v=3\).
In order to compare our results to the literature, we use Gaussian onsite potentials with zero mean and variance \(W\) for the small-world network and a uniform distribution \( \varepsilon_n \in [-W/2, W/2]\) for the RRG.
The random graphs were generated by the NetworkX library~\cite{hadberg2008exploring}. 
Examples of random graphs that we considered are shown in Fig~\ref{fig:random_graphs}.

The exact phase diagrams of these models continue to be a subject of intense debate.
For the case of the Anderson model on the RRG
it has been unequivocally established that for sufficiently strong disorder, \(W_L \approx 18.17\), Anderson localization occurs.
Several works have reported the existence of a non-ergodic extended (multi-fractal) phase for intermediate disorder strengths~\cite{deluca2014anderson, altshuler2016nonergodic, kravtsov2018non} as sketched in panel d) of Fig.~\ref{fig:schematic_phase_diagrams}, while others~\cite{tikhonov2016anderson, tikhonov2019critical,  biroli2022critical} argue that any non-ergodic behaviour is due to strong finite-size effects, whose scale diverges exponentially at both sides of the transition.
It is worth mentioning also the supposition stated in Ref.~\cite{pino2020scaling} that there is no ergodic but only a fractal phase.
For the small world networks the critical properties depend on two length scales, that diverge with different critical exponents which leads to strong finite size effects~\cite{garcia-mata2017scaling, garcia-mata2020two, garcia-mata2022critical}.
Here we are interested in relatively small system sizes where all studies agree that there is a range of disorder strength exhibiting a regime of multi-fractal states.

\begin{figure}
    \includegraphics[width=0.75\columnwidth]{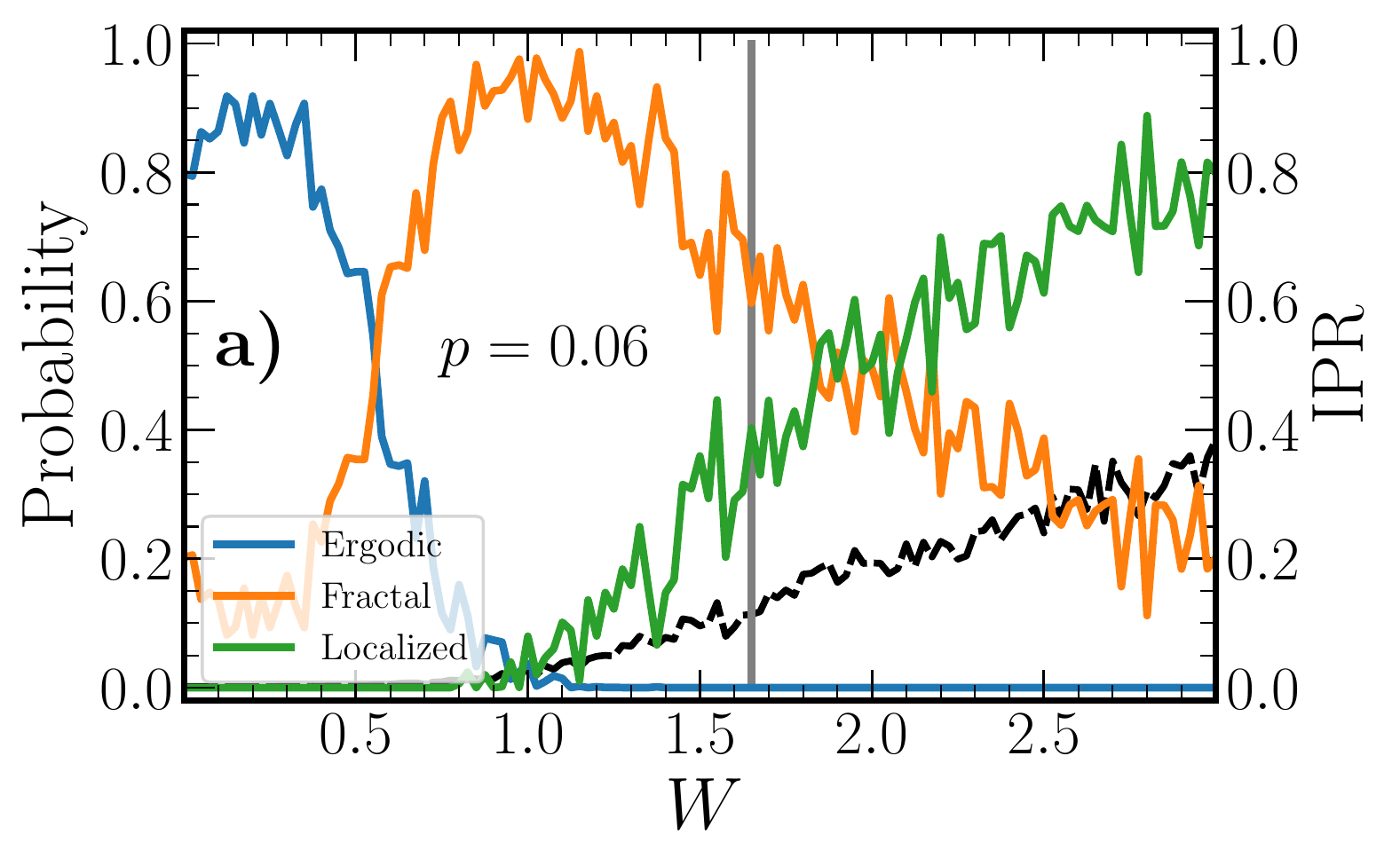}
    \includegraphics[width=0.75\columnwidth]{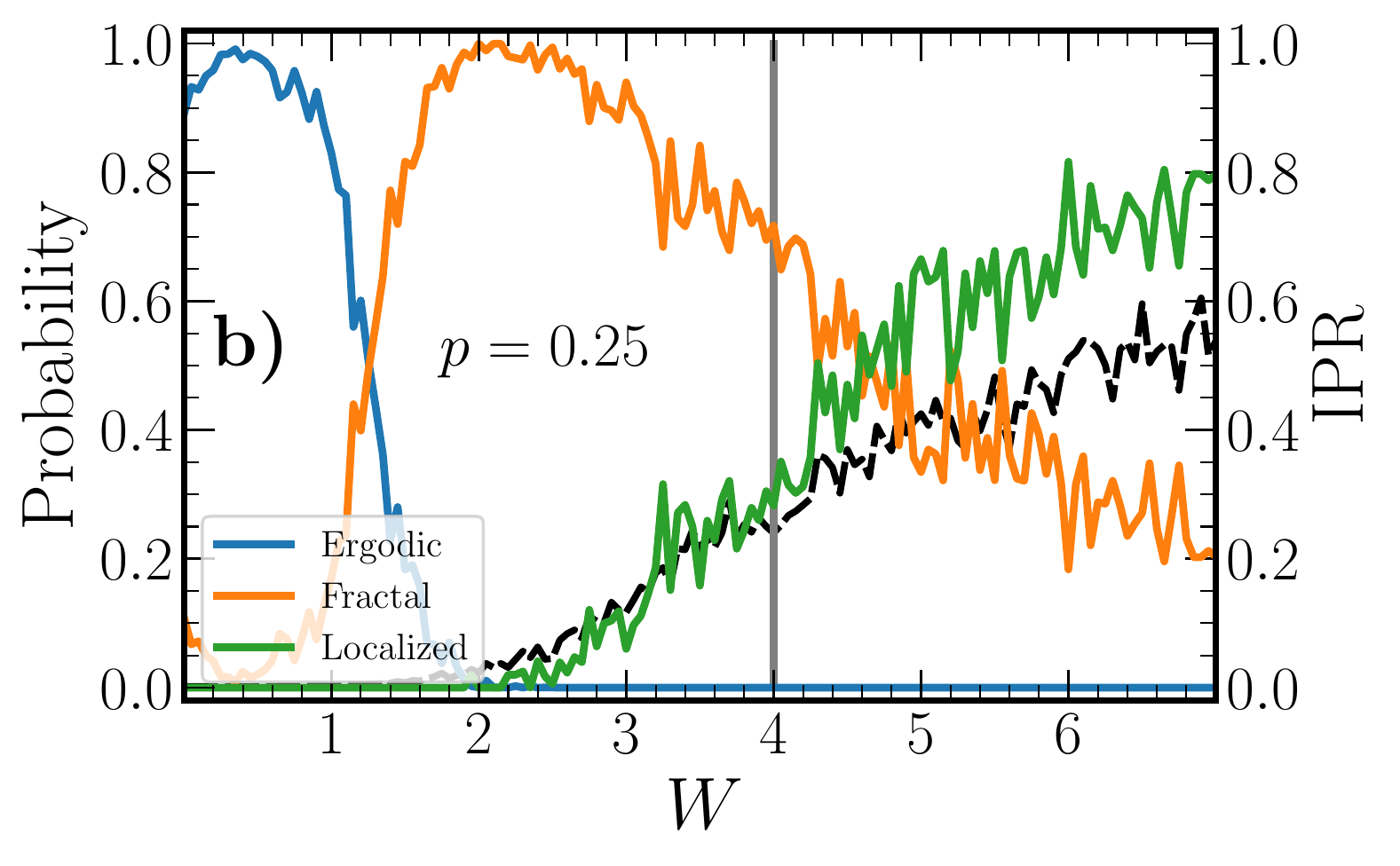}
    \includegraphics[width=0.75\columnwidth]{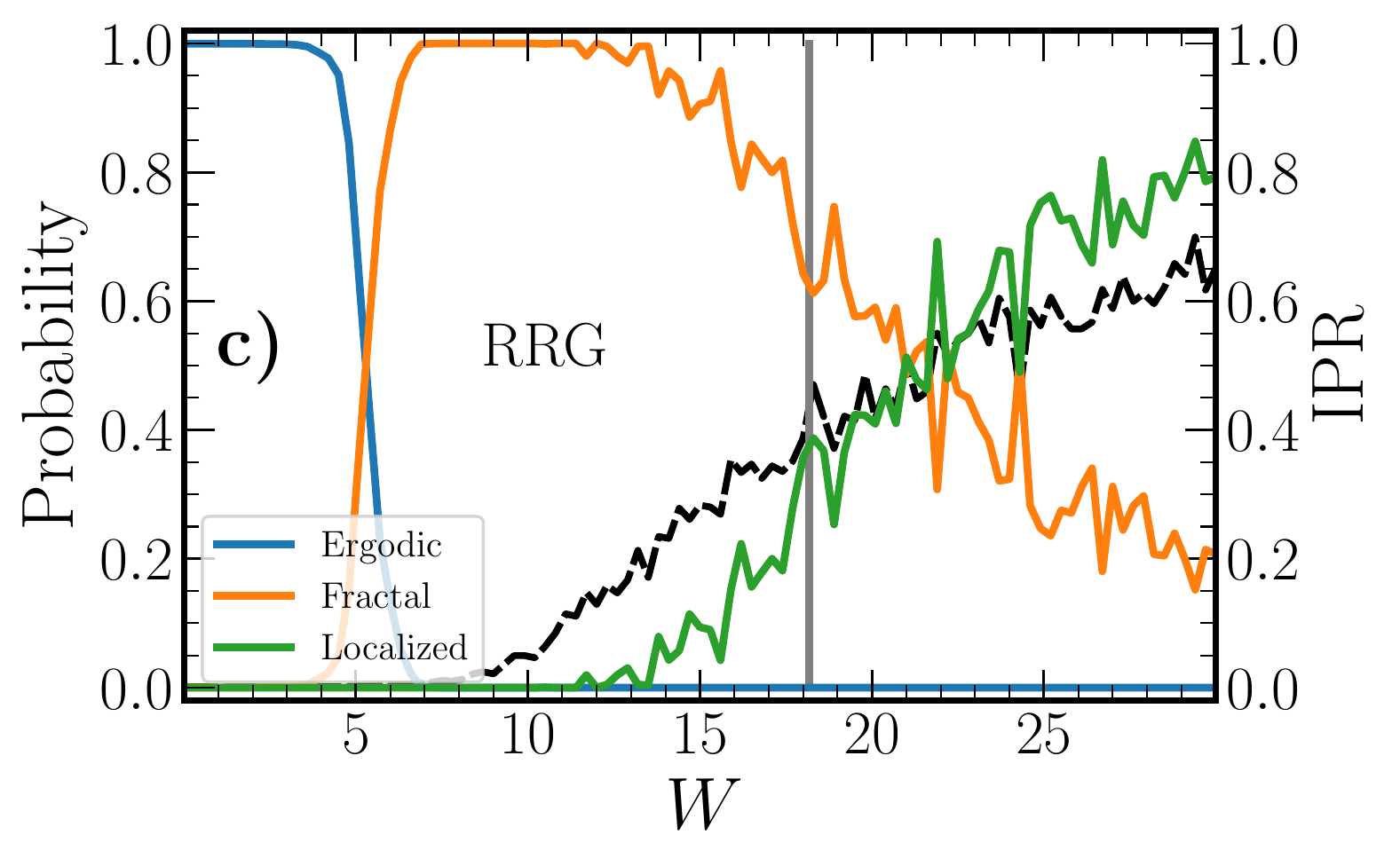}
    \caption{
        Testing the trained CNN for the Anderson model on the small-world network with \(p = 0.06, 0.25\) and the RRG are shown in panels a), b) and c), respectively.
        The probabilities of each of the phases are shown as well as the average IPR (black dashed line) as a function of the onsite disorder strength \(W\). 
        The gray vertical line indicates the numerical value for the Anderson transition \(W_L\).
        %\alexei{AA: make the Anderson transition line thicker to be better visible}
    }
    \label{fig:AM_RRG_testing}
\end{figure}

In Fig.~\ref{fig:AM_RRG_testing} we show the results of testing the CNN with the wave functions of the Anderson model on random graphs.
We increase \(W\) through \([0, 3], [0, 7], [0, 30]\), in steps of \(0.025, 0.05, 0.3\) for panels a), b), c), respectively.
For each \(W\), we generate \(50\) realizations of the disorder. 
For each realization we input the wave function with energy closest to \(\epsilon=0\) to the CNN. 
We then average the probabilities output by the CNN over the \(50\) realizations.
The values of the disorder where the Anderson transition occurs are given by \(W_L \approx 1.65, 4.0\) for the small world networks~\cite{garcia-mata2022critical} and \(W_L \approx 18.17\) for the RRG. 
These values are indicated by gray vertical lines. 
The CNN identifies ergodic, fractal and localized phases.
Note the similarity of the results obtained for different random graphs.
The ergodic transition is found at small disorder strengths of about \(W \sim 0.3 \cdot W_L\), whereas the Anderson transition is blurred. 
Identifying the crossing point of the probabilities \(P_F\) and \(P_L\) with the Anderson localization transition yields that the CNN overestimates the value of \(W\) where it occurs.
There are several possible explanations for this discrepancy: 
i) the criterion \(P_F = P_L\) overestimates the transition, which has already been seen in the case of the AAH model, or 
ii) due to the finite size effects the multi-fractal regime extends further into the localized phase (see Fig. 14 of Ref.~\cite{garcia-mata2022critical}).

\subsubsection{Mass deformed Sachdev-Ye-Kitaev model \normalfont{(\textbf{SYK})}}

A modification of the Sachdev-Ye-Kitaev model~\cite{sachdev1993spin, maldacena2016remarks, polchinski2016the} was introduced recently, called \emph{mass-deformed SYK model}~\cite{garcia-garcia2018chaotic, kim2021comment, garcia-garcia2021reply, nosaka2018the, monteiro2021minimal, monteiro2021quantum}. 
It is a many-body model of an even number, \(N_M\), of interacting Majorana fermions \(\hc_i, \ i = 1, \dots, N_M\), which obey the Clifford algebra \(\left\{ \hc_i,\, \hc_j\right\} = \delta_{ij}\). 
The Hamiltonian comprises one-body and two-body parts, both being all-to-all connected,
\begin{align}
    \label{eq:mass_deformed_H}
    & \mh \equiv \frac{2}{\sqrt{N_M}}\mh_4 + \kappa \, \mh_2, \notag \\
    & \mh_4 = -\sum_{i<j<k<l} J_{ijkl} \hc_i \hc_j \hc_k \hc_l, \; \\
    & \mh_2 = i \sum_{i<j} J_{ij} \hc_i \hc_j \, , \notag
\end{align}
where the coupling constants \(J_{ijkl}\) and \(J_{ij}\) are Gaussian distributed with zero mean and variances \(6/N_M^3\) and \(1/N_M\), respectively. 
The Hamiltonian does not preserve the number of particles but their parity, yielding \(2^{N_M/2 - 1}\) for the dimension of the relevant Hilbert space.

The model is analytically solvable in the thermodynamic limit~\cite{monteiro2021minimal} and the analytical predictions have been verified numerically in finite systems~\cite{monteiro2021minimal, monteiro2021quantum, nandy2022delayed}. 
In Ref.~\onlinecite{monteiro2021minimal} the authors identified four regimes in the phase diagram of the model~\footnote{Note that our notation differs from that used in Ref.~\onlinecite{monteiro2021minimal}. 
It is the same as in Ref.~\onlinecite{nandy2022delayed}. 
The mapping between the two notations is given in Appendix A of Ref.~\onlinecite{nandy2022delayed}.} 
that can be distinguished by the localization properties of the eigenstates in the Fock space -- when studied in the eigenbasis of the one-body term.  
In regime I (\(\kappa < \kappa_1\)) the eigenstates are ergodically extended over the full Fock space. 
In regimes II (\(\kappa_1 < \kappa < \kappa_2\)) and III (\(\kappa_2 < \kappa < \kappa_3\)) the eigenstates are still ergodically extended, implying, in particular, that they do not have any fractal properties as extensively discussed in Ref.~\onlinecite{monteiro2021quantum}, 
but their extension is over energy shells whose dimension still scales exponentially in \(N\). 
In regime II (III), all (a fraction of) the nearest neighbors of a chosen unperturbed state are hybridized, respectively.
In regime IV (\(\kappa > \kappa_3\)) the eigenstates are localized in the Fock space. 
Interestingly, as a consequence of the one-body term being all-to-all, the eigenstates are fully delocalized in all regimes in the computational basis implying that localization properties \emph{must} be studied in the eigenbasis defined by the eigenstates of the one-body term; see Appendix~\ref{app:additional} for further details.
The values of the regime boundaries are 
\(\kappa_1 = \sqrt{(N_M-2)(N_M-3)/(2 N_M^3)}\), \(\kappa_2 = \sqrt{N_M} \, \kappa_1\) and \(\kappa_3 = Z/\sqrt{8 \rho} \, \mathcal{W}(2 \sqrt{\pi} Z)\), with \(\rho = \binom{N_M}{4}\), \(Z = \binom{N_M/2}{4}\) and \(\mathcal{W}(x)\) denoting the Lambert function. 
The schematic phase diagram is shown in panel e) of Fig.~\ref{fig:schematic_phase_diagrams}.

\begin{figure}
    \includegraphics[width=0.75\columnwidth]{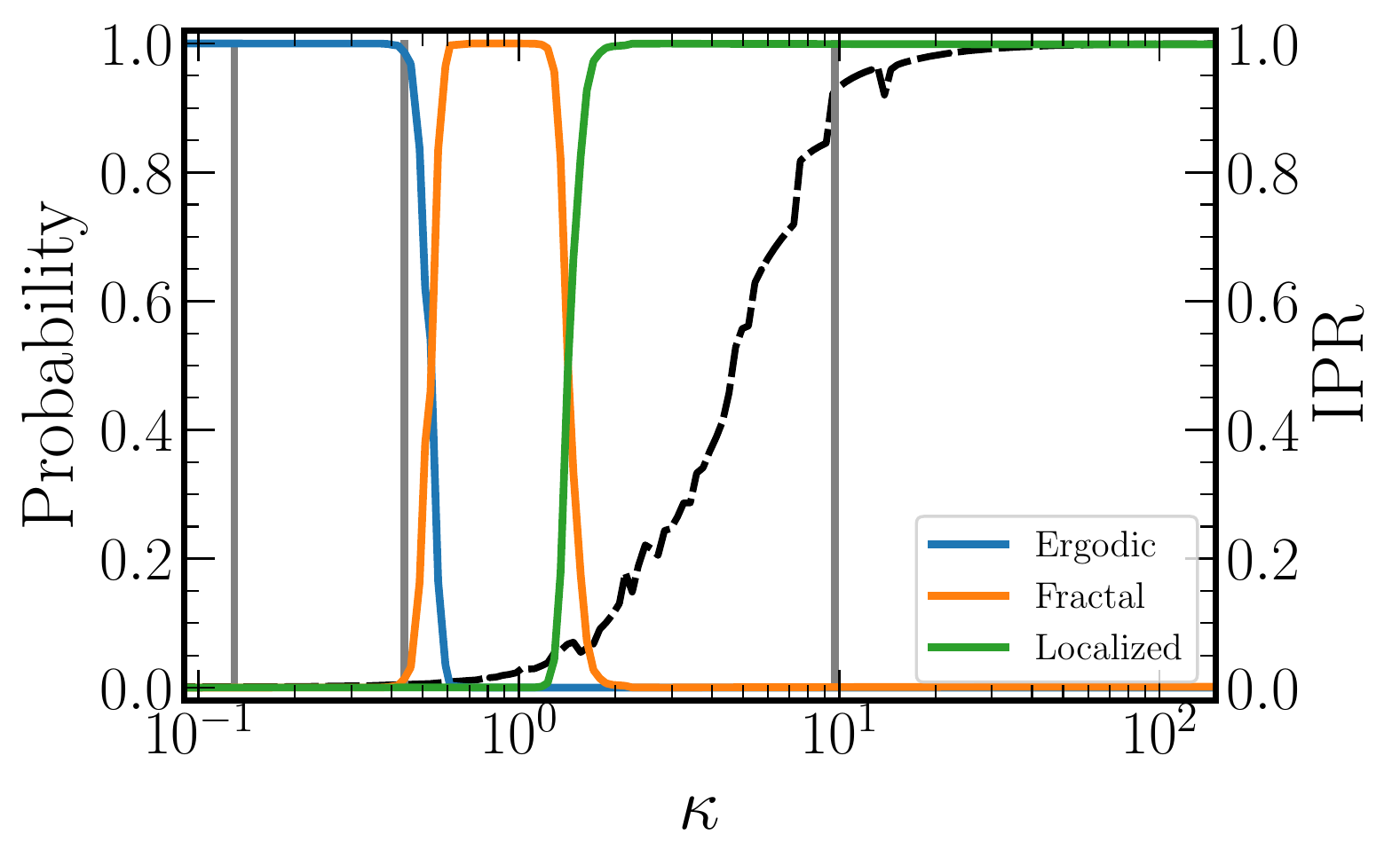}
    \caption{
        Testing the trained CNN on the mass-deformed SYK model.
        The probabilities of each of the phases are shown as well as the average IPR (dash black line) as a function of the parameter \(\kappa\), which is plotted on a logarithmic scale. 
        The gray vertical lines indicate the analytical values \(\kappa_1, \kappa_2, \kappa_3\) for the transitions.
    }
    \label{fig:SYK_testing}
\end{figure}

We test the CNN on \(5\) eigenfunctions expanded in the eigenbasis of \(\mh_2\) for each value of \(\kappa\in [0.1, 1000]\) on a logarithmic scale for 200 values of \(\kappa\). 
As can be seen in Fig.~\ref{fig:SYK_testing}, the CNN identifies three distinct phases, ergodic, fractal, and localized.
It should be stressed once again, that the model \emph{does not} have a genuine fractal phase~\cite{monteiro2021quantum}.
Nevertheless it is reasonable to expect that the CNN, once required to classify the intermediate regime and knowing only about the existence of three phases, finds highest similarity of it with the previously learned fractal phase of the gRP model.
More quantitatively, the ergodic transition is found at \(\kappa_E \approx 0.5\), which is slightly larger than the value of \(\kappa_2\), while the Anderson transition is observed at \(\kappa_L \sim 2\) which underestimates the analytical value. 
However, it is interesting to note that the values \(\kappa\) at the transitions, \(\kappa_E,\) and \(\kappa_L\) identified by the CNN coincide well with those where another quantity of interest in the study of many-body systems, namely
the \emph{adiabatic gauge potential}, which is related to the \emph{fidelity susceptibility}~\cite{sierant2019fidelity, maksymov2019energy, pandey2020adiabatic, sels2021dynamical}, shows maximally chaotic behaviour~\cite{nandy2022delayed}, 
the onset being close to \(\kappa_E\), whereas it exhibits a peak at \(\kappa_L\). 
Similar behavior is observed in the gRP model~\cite{skvortsov2022sensitivity, cadez2023gRP}.

\section{Discussion and conclusions}
\label{sec:conclusions}

In this work we have trained a CNN to identify ergodic, fractal and localized states in the gRP model. 
The model was chosen since its phase diagram is known exactly~\cite{kravtsov2015random}.
The main result of this work is a demonstration of the generalization capability -- the same network was applied to diverse systems: single-particle models, including random-matrix models, random graphs, and a many-body quantum system. 
The training set of only \(500\) states per class and testing sets of several states per test point suffice to discern the three phases for a fixed system size, making the network very efficient. 
Thus, it can be effectively used also for full spectra.

On the other hand we have found that improving the precision is not straightforward. 
For example using larger training sets (\(5000\) states for each class in the training) or larger system sizes (not shown) produces results similar to the ones presented. 
Thus more detailed analyses such as finite-size scaling are hard to perform. 
A similar conclusion was reported recently in an attempt to use a neural network for the detection of the many-body localization transition~\cite{theveniaut2019neural}. 

Note that in our approach we use relatively small system sizes, implying that it may exhibit strong finite-size effects. 
This is particularly evident in the cases of the 3DA model, the PLBM models, and the Anderson model on random graphs. 
Yet, comparable sizes are frequently used in numerical simulations, so especially for these the results deduced from the CNN provide relevant information on wave function localization properties. 
Furthermore, in all these models the characteristic length scales corresponding to the three phases becomes large in a non-negligible range around the parameter values where the transition takes place, 
implying that there the interpretation of phases or locating the transition can be complicated like, e.g., for the Anderson model on an RRG. 
Nevertheless, our approach provides a good qualitative agreement with other procedures as outlined in this work.

During the completion of this work, we became aware of a recent work~\cite{nelson2023phase}, where the authors use an artificial neural network to study delocalized, multifractal and localized phases in a variant of the AAH model, the long-range AAH model. 
They construct a multi-layer perceptron, which is a dense neural network that is trained on the long-range AAH and tested on the AAH and the training model itself, implying that our study of generalization capabilities is obviously more extensive.
Another disadvantage as compared to the gRP model is, that in the long-range AAH the multifractal or localized phases coexist with the delocalized one. 
This is a drawback for preparing high-quality training sets. 
Above all, in distinction to the AAH, the gRP model is an all-to-all model, which thus can be applied to various types of systems, including single-particle systems like the AAH, as demonstrated in this work. 

\section*{Acknowledgments} 
\label{sec:acknowledgments}

AA, BD, DR and T\v{C} acknowledge financial support from the Institute for Basic Science (IBS) in the Republic of Korea through the project IBS-R024-D1.
TO and KS were supported by JSPS KAKENHI Grants No. 19H00658.
TO was supported by JSPS KAKENHI Grants No. 22H05114.

\appendix

\section{CNN hyperparameters}
\label{app:CNN}

As described in Section~\ref{subsec:CNN_training} we use a simple CNN which we constructed using Keras~\cite{chollet2015keras} as the frontend and TensorFlow~\cite{tensorflow2015whitepaper} as the backend. 
The CNN parameters related to its structure, such as the number of different layers, their sizes, etc. are known as \emph{hyperparameters}. 
We have tried several CNN architectures and found that the one given in Table~\ref{tb:hyperparameters} performed best. 
The convolutional layers apply filtering on the data using a number of \emph{kernel}s of given sizes. 
The number of kernels that have been employed is provided in the table under filters. 
The stride is a translation for which the kernel moves along the data. 
The output size corresponds to that of the data in each of the layers. 
In the pooling layers we have used max pooling. 
No padding was used.
To minimize the categorical crossentropy we have used the Adam optimizer.

\begin{table}
    \begin{center}
        \begin{tabular}{l|c|c|c|c} 
            layer class & filters & kernel size & stride & output size\\
            \hline \hline
            input  & & & & (1, 2048)\\ 
            convolutional 1 & 64 & 256 & 1 & (64, 1793) \\
	        pooling 1       &    &   2 & 2 & (64,  896) \\
            convolutional 2 & 16 & 128 & 1 & (16,  769) \\
            pooling 2       &    &   2 & 2 & (16,  384) \\
            dense 1         &    &     &   & (16) \\
            dense 2         &    &     &   & (3) \\
        \end{tabular}
    \end{center}
    \caption{
        The structure and hyperparameters of the CNN used. 
        The total number of trainable parameters is \(245808\).
    }
    \label{tb:hyperparameters}
\end{table}

\section{Examples of wave functions}
\label{app:wavefunctions}

\begin{figure}
    \includegraphics[width=0.99\columnwidth]%{figures/wavefunctions_N2048.pdf}
    {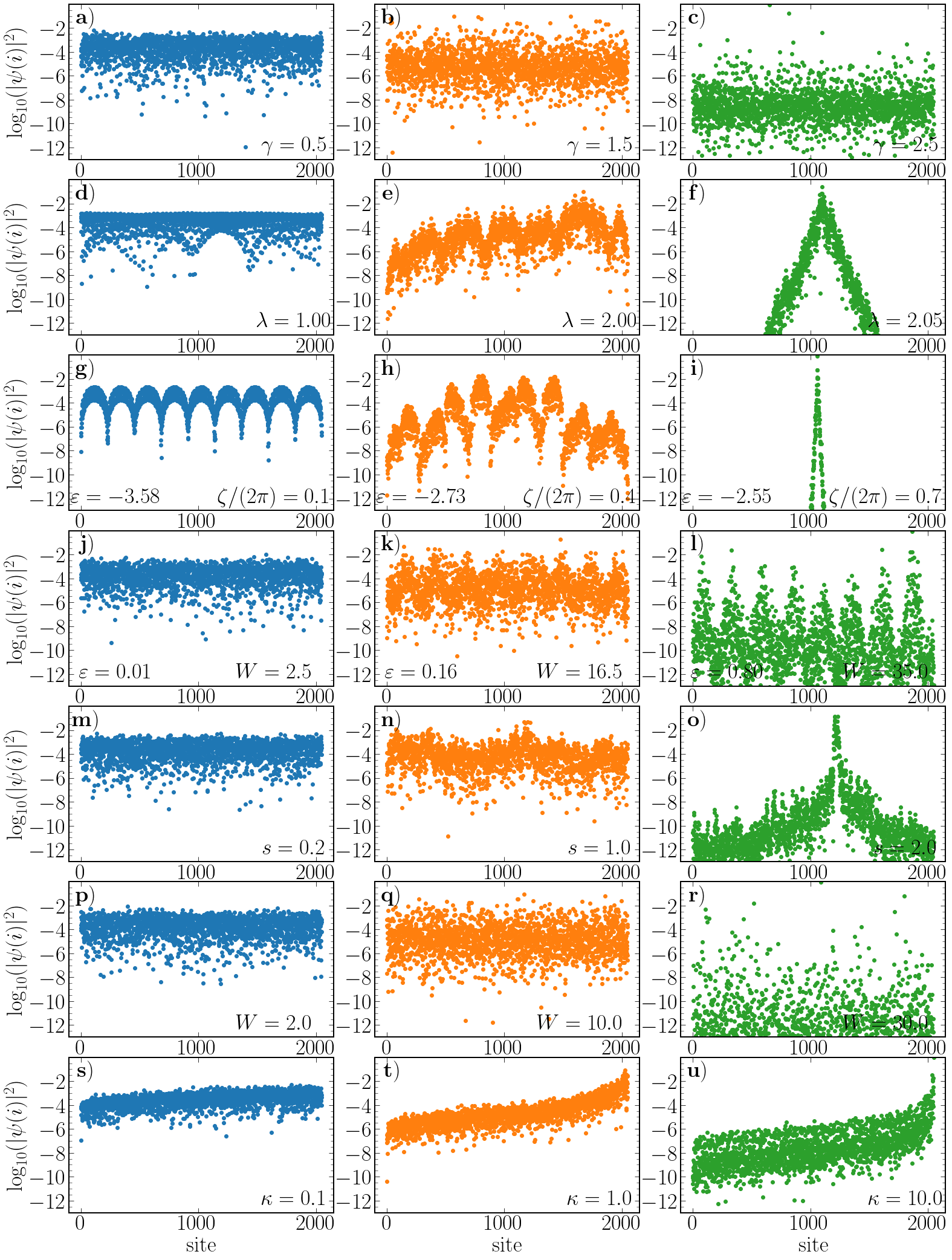}
    \caption{
        Examples for the eigenvector occupations for 
        a)-c) the gRP model, 
        d)-f) the AAH model, 
        g)-i) the eH model, 
        j)-l) the 3DA model, 
        m)-o) the PLBM model, 
        p)-r) the Anderson model on RRG and 
        s)-u) the mass deformed SYK model. 
        Note the logscale for the y axis. 
        The values of the parameters used are given in panels. See main text for details.
    }
    \label{fig:wavefunctions}
\end{figure}

Examples of the eigenvector occupations - the raw data that we use as input for CNN - are shown in Fig.~\ref{fig:wavefunctions}. 
Note that these are the original states that are not centered. 
The left, middle and right panels display states that were classified by the CNN as extended, fractal and localized, respectively. 
Results are shown for a)-c) the gRP model, d)-f) the AAH model, g)-i) the eH model, j)-l) the 3DA model, m)-o) the PLBM model, p)-r) the Anderson model on RRGs and s)-u) the mass deformed SYK model. 
For the gRP, PLBM and Anderson model on RRGs we used the computational basis. 
In the single-particle models (AAH, eH and 3DA) the computational basis is the real space basis. 
For the mass deformed SYK model we use the eigenbasis of \(\mh_2\). 
The probability distributions of the gRP and PLBM were studied recently in Refs.~\onlinecite{bogomolny2018eigenfunction} and~\onlinecite{bogomolny2018power}. 
In the ergodic regime they are given by the Porter-Thomas distribution~\cite{porter1965statistical,mehta2004random,haake2018quantum}. 
For the eH and 3DA model we show states from the lowest band (see Fig.~\ref{fig:extended_Harper}) and from the vicinity of band center, respectively.

\section{Additional results}
\label{app:additional}

Here we show some additional results that address three points, 
(i) larger testing data set, 
(ii) larger training data set and 
(iii) the importance of the choice of basis. 

\begin{figure}
\includegraphics[width=0.49\columnwidth]{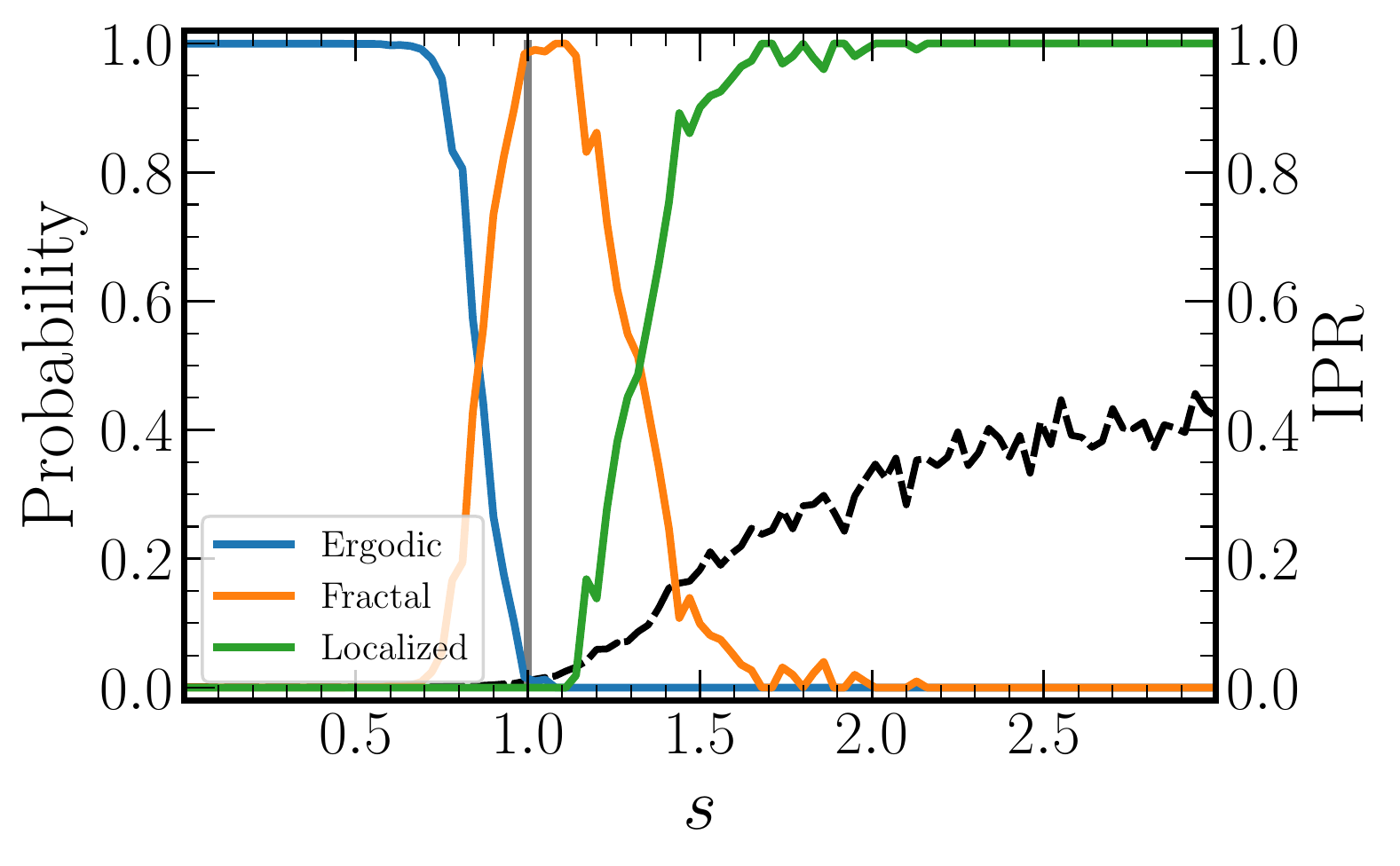}
\includegraphics[width=0.49\columnwidth]{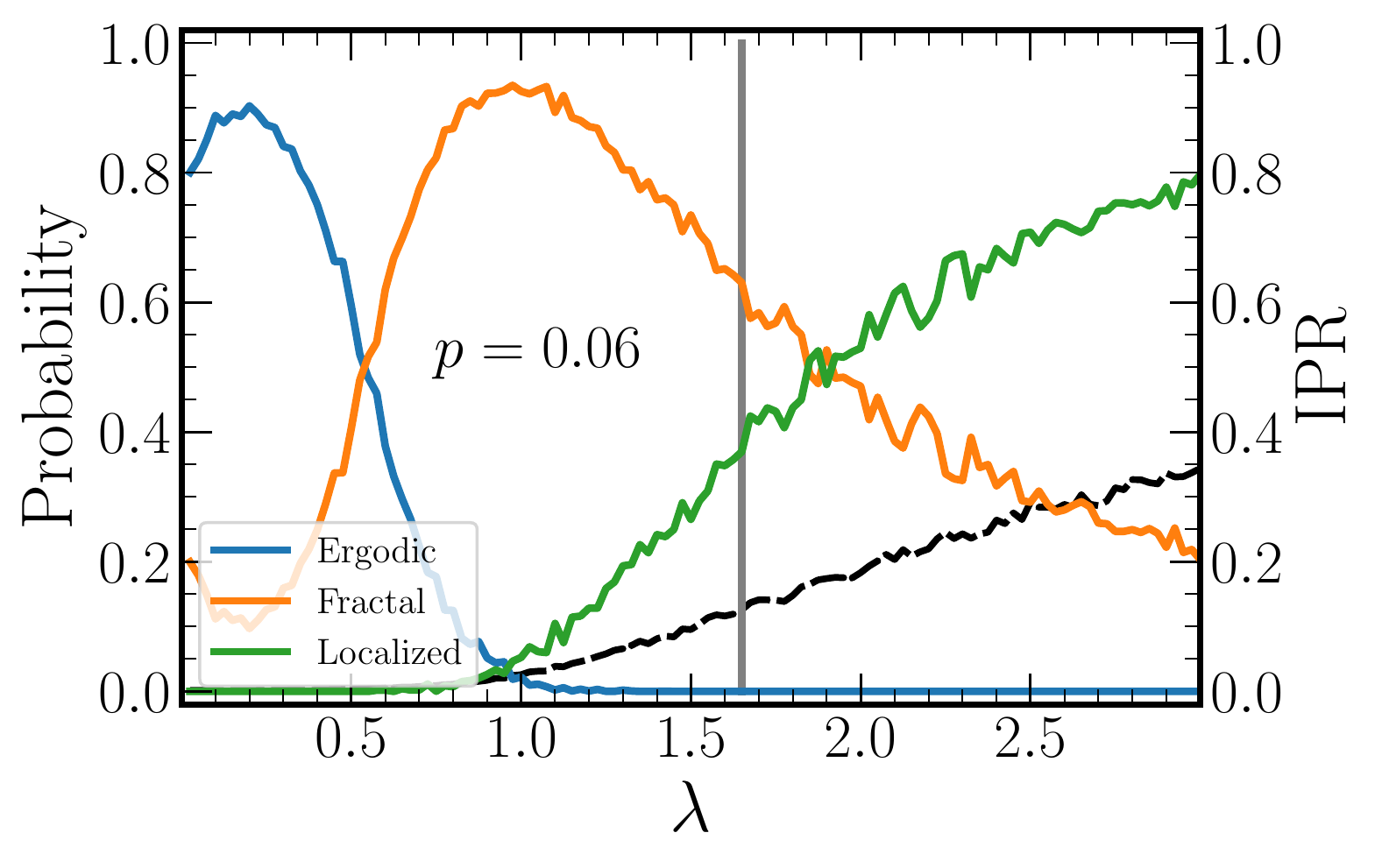}
    \caption{
        Generalization capability on the PLBM (left panel) and the Anderson model on a small-world network with \(p = 0.06\) (right panel) on a larger data set reduces fluctuations. 
        We used \(50\) and \(500\) realizations per each point for left and right panel, respectively.
    }
    \label{fig:larger_testing_set}
\end{figure}

Especially in the cases of the PLBM and the Anderson model on random graphs the testing data averaged over \(5\) and \(50\) states, as shown in Figs.~\ref{fig:PLBM_testing} and \ref{fig:AM_RRG_testing}, respectively, fluctuate strongly. 
In Fig.~\ref{fig:larger_testing_set} we demonstrate that increasing the testing data amount by a factor of \(10\) reduces the fluctuations considerably, however the transition values of the associated parameter do not change up to currently achievable precision.

\begin{figure*}
    \includegraphics[width=0.19\textwidth]{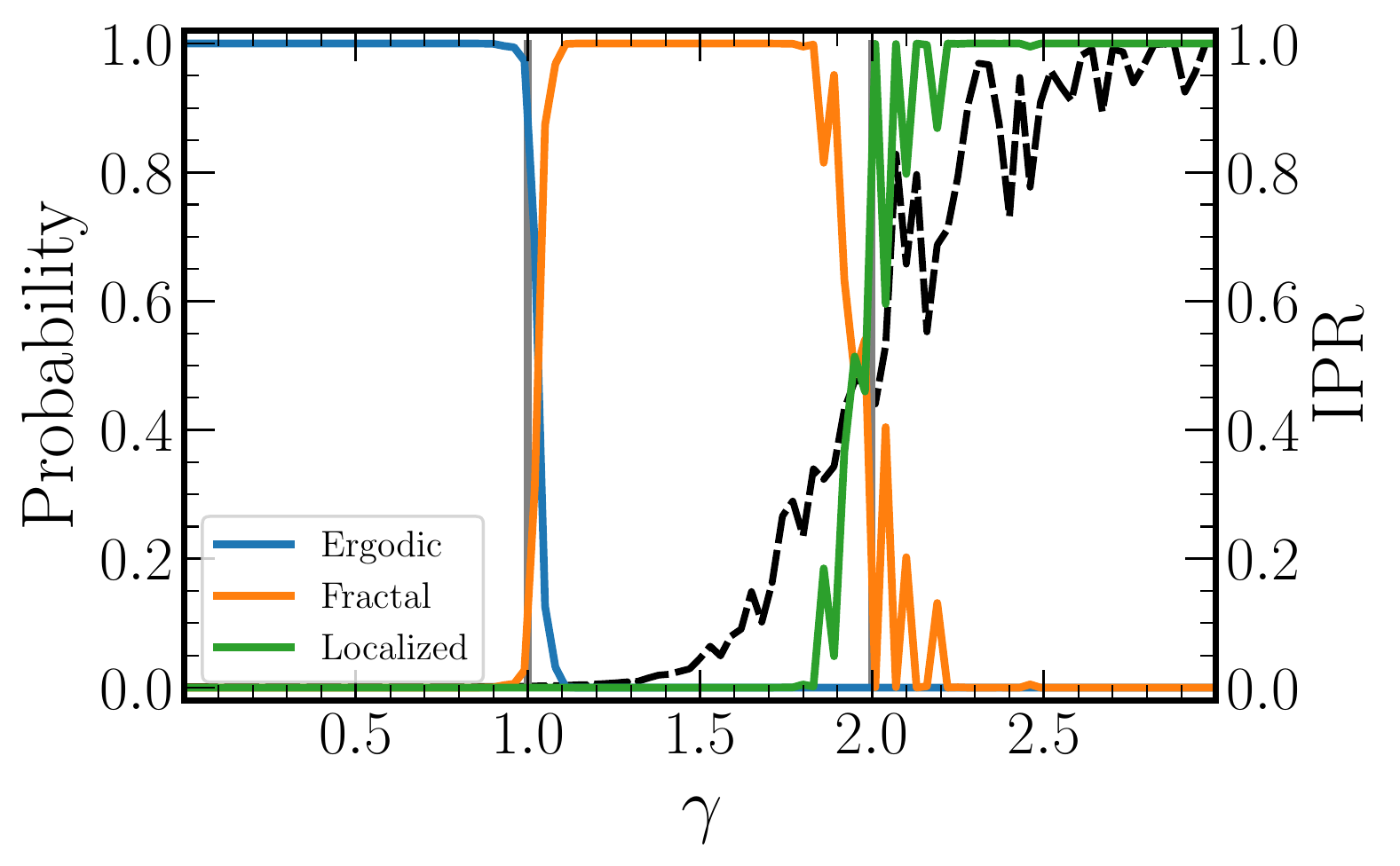}
    \includegraphics[width=0.19\textwidth]{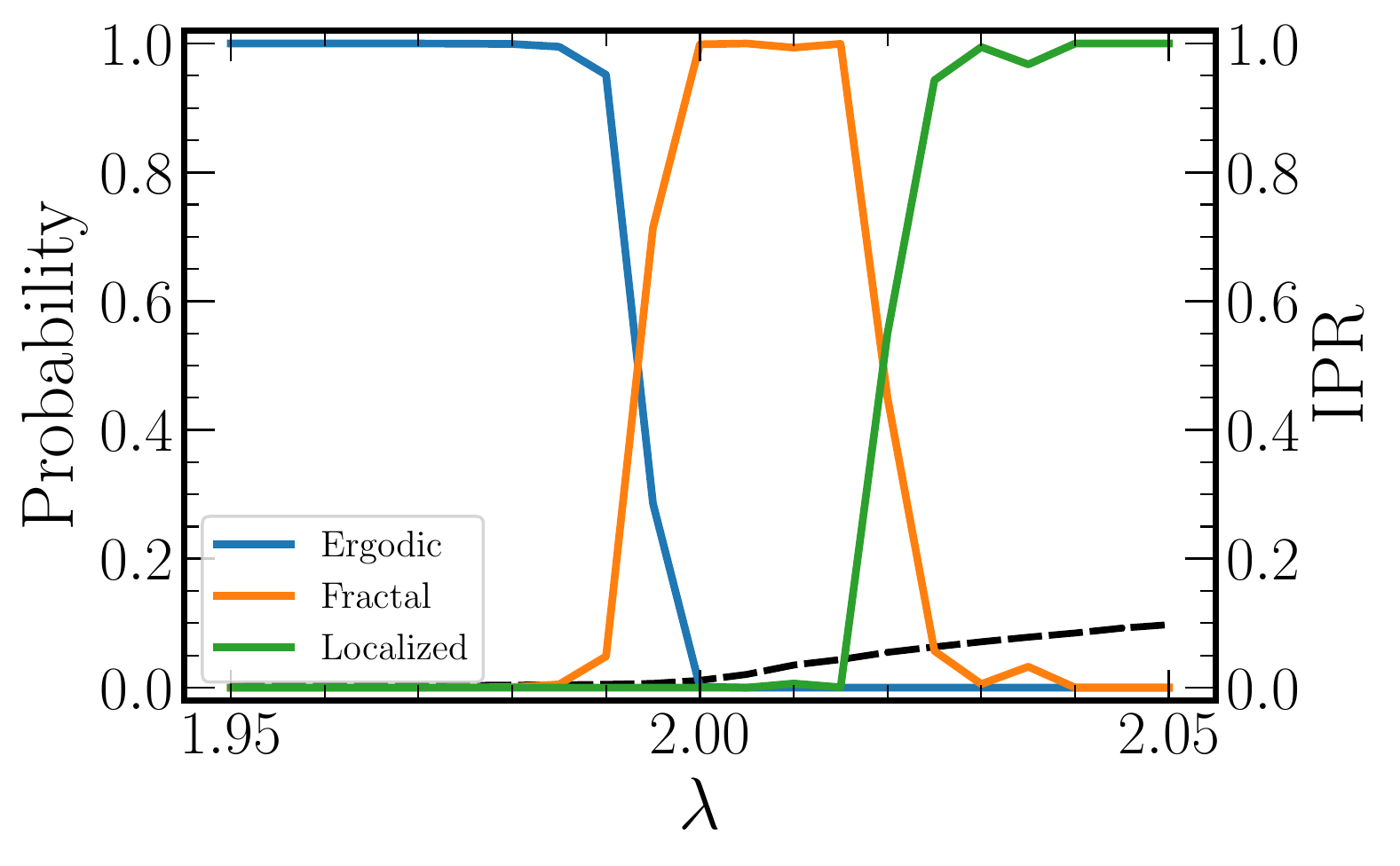}
    \includegraphics[width=0.19\textwidth]{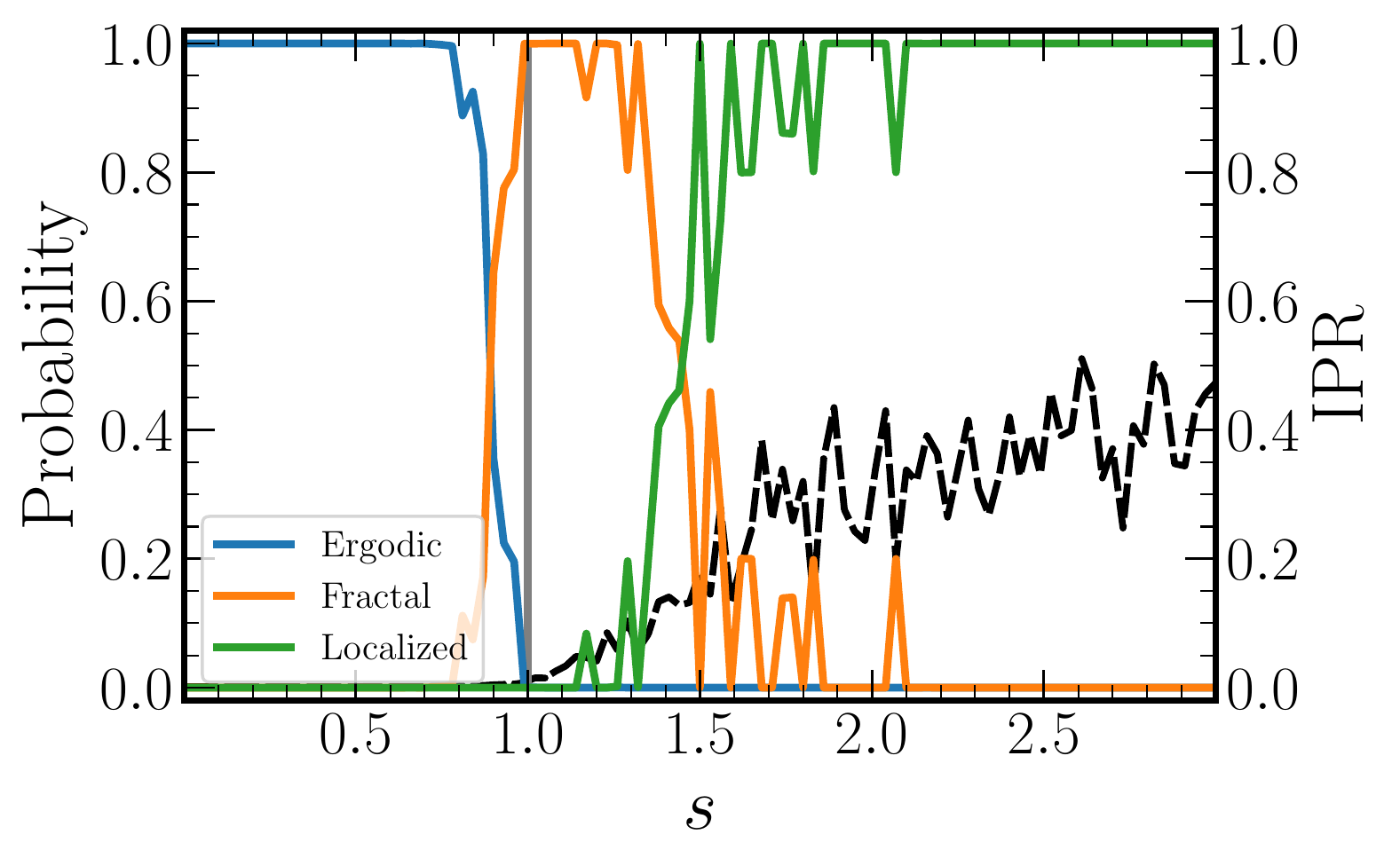}
    \includegraphics[width=0.19\textwidth]{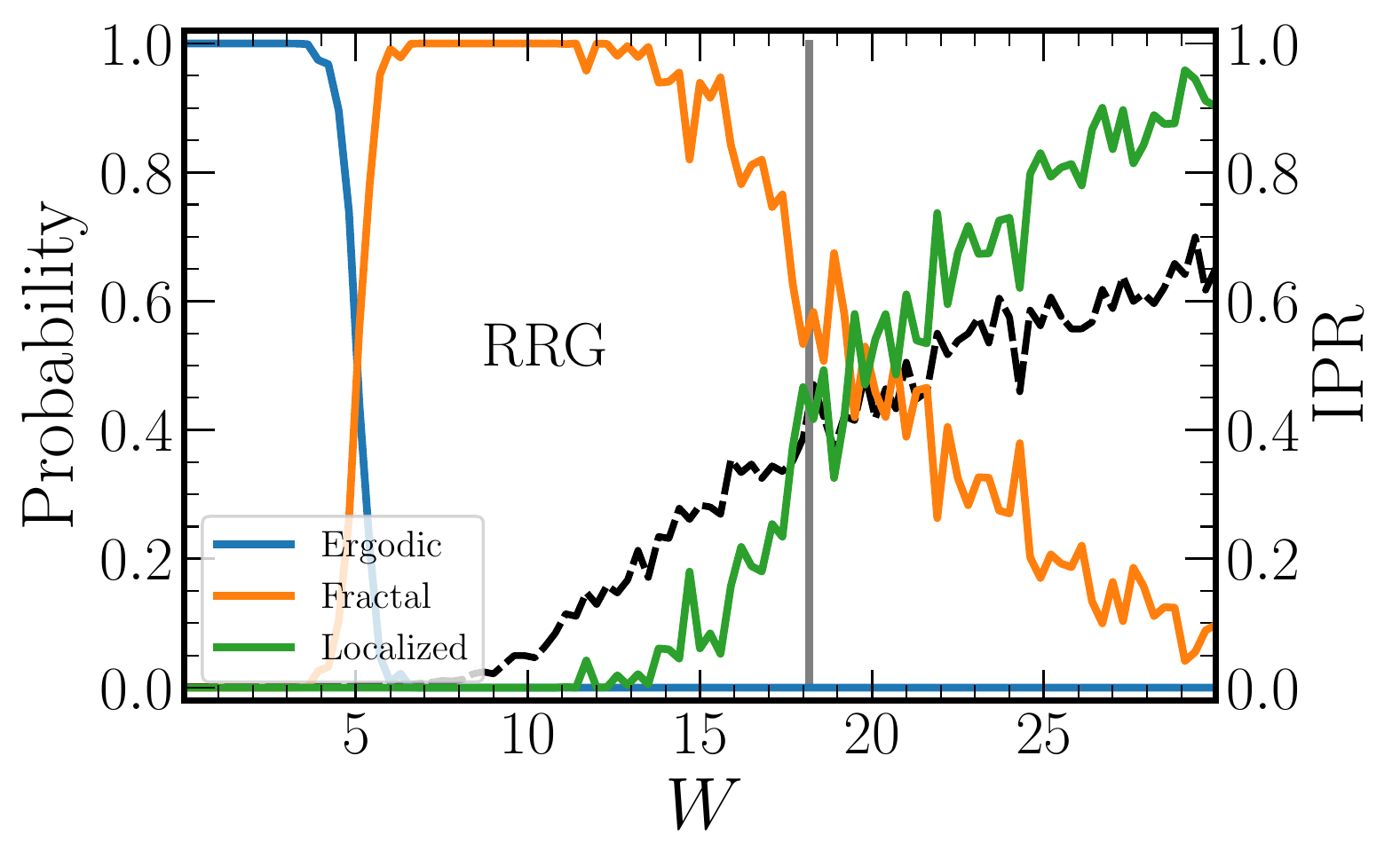}
    \includegraphics[width=0.19\textwidth]{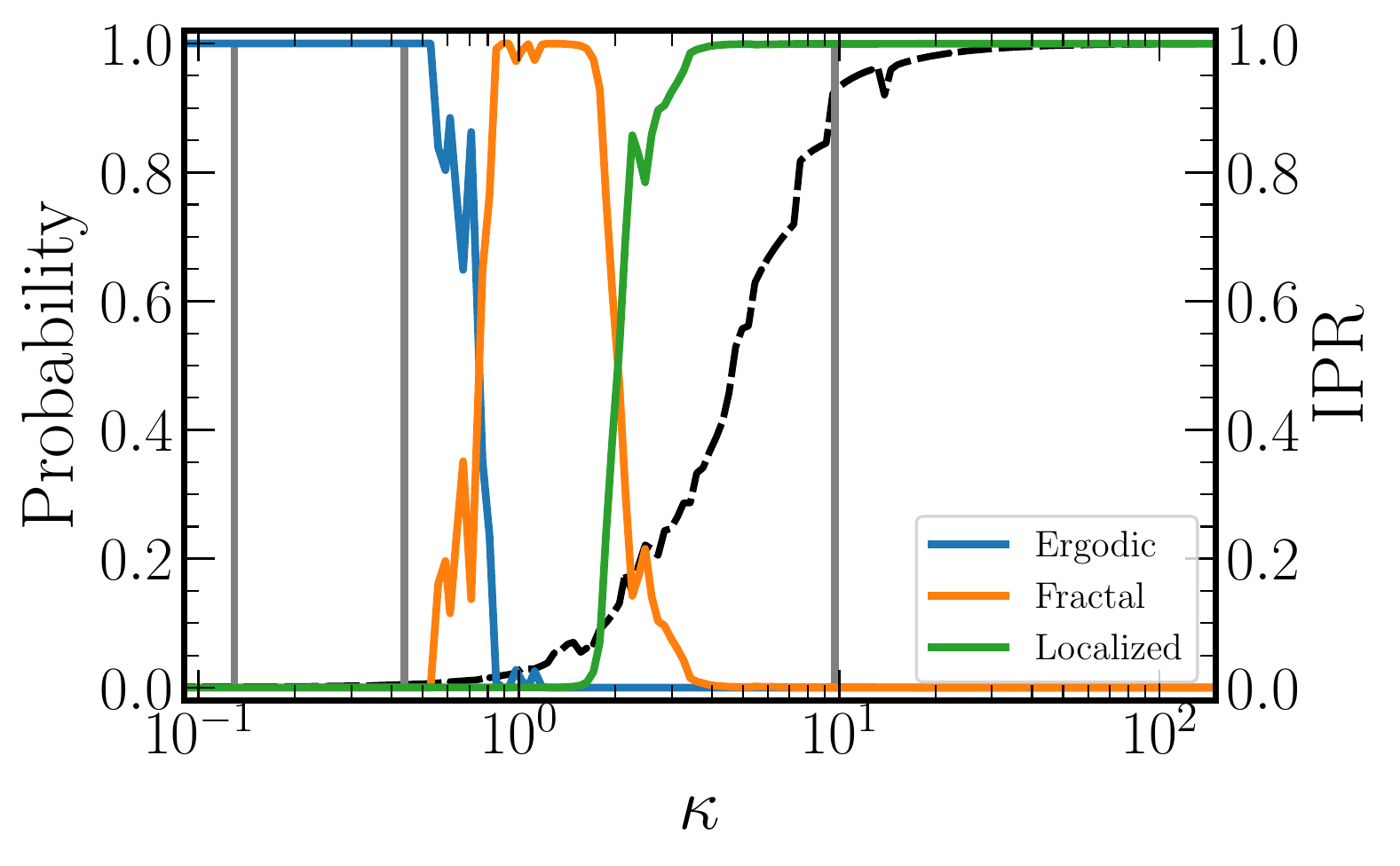}
    \caption{
        Training the CNN on a larger data set (5000 eigenfunctions per phase) slightly improves testing results on the same model, as seen in the leftmost panel. 
        However the generalization capability does not necessarily improve as can be seen in the cases of (from left to right): AAH model, PLBM, Anderson model on RRG and the mass deformed SYK model. 
        The same testing data as in the main text was used.
    }
    \label{fig:larger_training_set}
\end{figure*}

Regarding the supervised machine learning the amount of training data can be important since it can improve the performance of the trained network. 
In the main text we have trained the network on only \(500\) images per phase. 
Here we used the same architecture as given in Appendix~\ref{app:CNN} and trained the network on \(10\) times more data, again based on the gRP model. 
The results of testing are shown in Fig.~\ref{fig:larger_training_set}. 
We can see that the performance of the CNN can improve slightly when tested on the same gRP model, however the transfer learning generalization capability does not seem to improve. 

\begin{figure}
    \includegraphics[width=0.7\columnwidth]{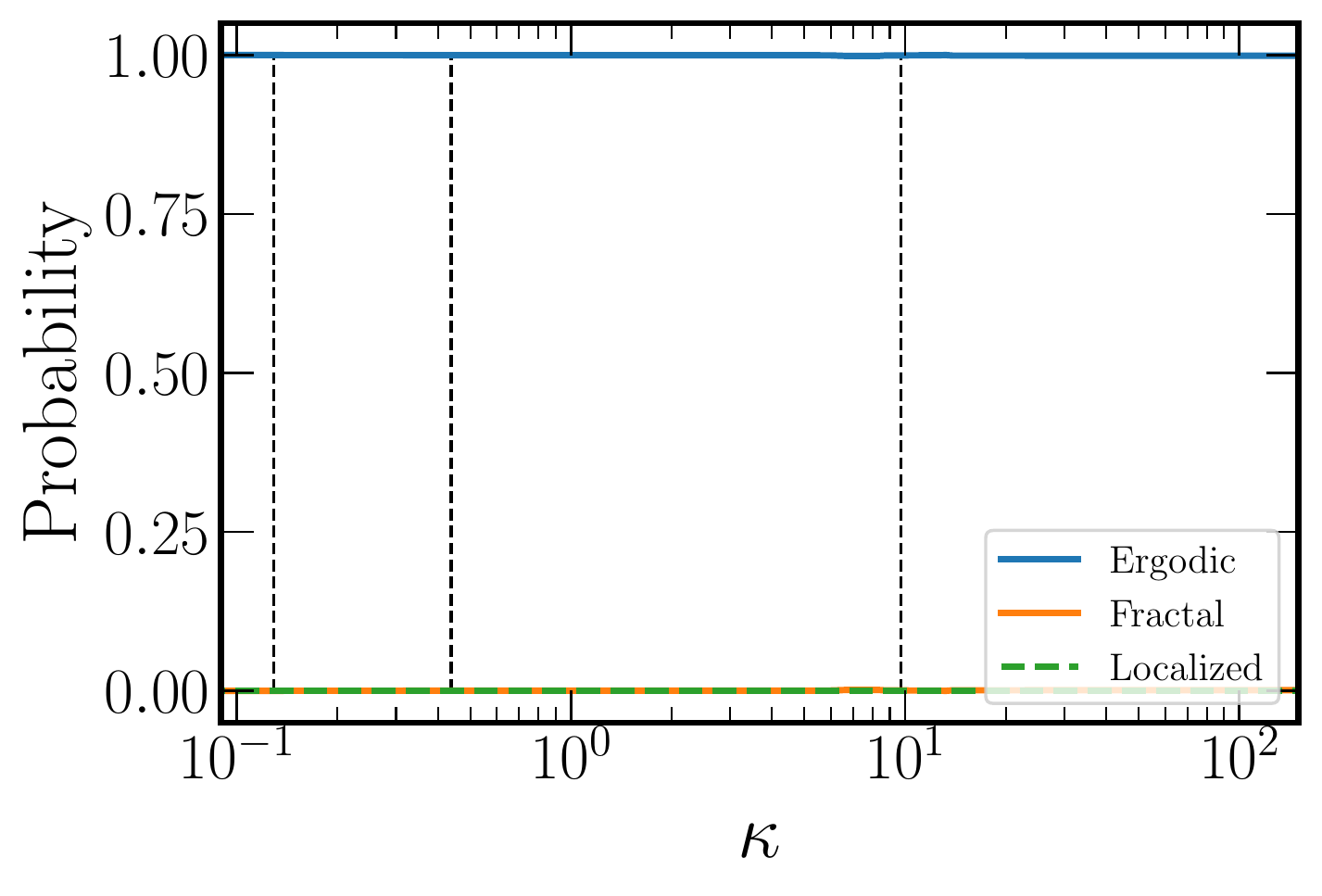}
    \caption{
        Testing results of CNN when applied to data from the computational basis of the mass deformed SYK model, showing complete delocalization.
    }
    \label{fig:delocalized_SYK}
\end{figure}

Finally we note that the localization properties depend crucially on the choice of basis. 
We show exemplary for the mass deformed SYK model in Fig.~\ref{fig:delocalized_SYK} the CNN testing results when the chosen basis is the computational basis of the full Hamiltonian Eq.~\eqref{eq:mass_deformed_H}, rather than the eigenbasis of \(\mh_2\).

\section{Criteria for ergodicity and localization for full random matrices}
\label{app:criteria}

The rule of thumb criteria for ergodicity and localization in dense matrices are based on the following sums of the averages of matrix elements~\cite{bogomolny2018power, khaymovich2020fragile}
\begin{align}
    \label{eq:S_q}
    S_q(N) = \frac{1}{N A^q} \sum_{n,m = 1}^N \langle |H_{nm}|^q \rangle,
\end{align}
where \(q = 1,2\), \(A = \sqrt{\langle |H_{nn}|^2 \rangle} = \sigma_d\) and \(N\) is the dimension of the matrix. 
The criteria are:
\begin{itemize}
    \item The Anderson localization criterion states that when \(\lim_{N \to \infty} S_1(N) < \infty\) the eigenstates are localized.
    \item The ergodicity criterion states that when \(\lim_{N \to \infty} S_2(N) \to \infty\) the eigenstates are ergodic.
    %\item In the case that \(\lim_{N \to \infty} S_1(N) \to \infty\) and \(\lim_{N \to \infty} S_2(N) < \infty\) the states are extended but non-ergodic. 
    \item Additionally a sufficient condition for full ergodicity is met~\cite{khaymovich2020fragile} if \(\lim_{N \to \infty} S_1(N) \to \infty\), \(\lim_{N \to \infty} S_2(N) \to \infty\) and \(\lim_{N \to \infty} \bar{S}(N) \to \infty\), where 
    \begin{align}
        \label{eq:barS}
        \bar{S}(N) = \frac{\bigl(\sum_{m} \langle |H_{nm}|^2 \rangle_{\mathrm{t}}\bigr)^2}{A^2 \, S_2(N)},
    \end{align}
    and the typical value is given as \( \langle |H_{nm}|^2 \rangle_{\mathrm{t}} = \exp\langle \ln(|H_{nm}|^2)\rangle \). 
\end{itemize}

For a Gaussian distribution given by
\begin{align}
    {\cal{P}}(x) = \exp{-x^2/(2 \sigma^2)}/\sqrt{2 \pi \sigma^2}
\end{align}
the moments can be calculated exactly \(\langle |x|^q \rangle = 2^{q/2} \sigma^q \Gamma{(q/2+1/2)/\sqrt{\pi}}\) for \(q > -1\), with the Gamma function \(\Gamma(a) = \int_0^{\infty} x^{a-1} \exp(-x) \mathrm{d}x\), 
while the typical value of the second moment is \(\langle |x|^2 \rangle_{\mathrm{t}} = \sigma^2/\bigl[2 \exp(\gamma_E) \bigr] \), with \(\gamma_E\) being the Euler-Mascheroni constant. 
Note that since \(\bar{S} = S_2/\bigl[2 \exp(\gamma_E)\bigr]\)
for Gaussian distributions, the criterion for full ergodicity coincides with the criterion for ergodicity.

For the gRP model, the variances are given in Eq.~\eqref{eq:gRP_variances} and we get \(S_1(N) =  \sqrt{2/\pi} \bigl[ 1 + 1/\sqrt{2} (N-1) N^{-\gamma/2}\bigr]\) and \( S_2(N) = 1 + 1/2 (N-1) N^{-\gamma}\). 
Taking the limit \(N \to \infty\) and using the above criteria follows the phase diagram of gRP. 

For the PLBM, the variances are given in Eq.~\eqref{eq:PLBM_variances} and we get \(S_1(N) =  \sqrt{2/\pi} \bigl[ 1 + 1/\sqrt{2} \, H_{N,s}\bigr]\) and \( S_2(N) = 1 + 1/2 \, H_{N,2s}\), where \(H_{N,s}\) are the generalized harmonic numbers. 
For \(s > 1\) we have \(\lim_{N\to\infty}H_{N,s} = \zeta(s)\), where \(\zeta(s)\) is the Riemann zeta function, whereas for \( 0 < s \le 1\) the limit can be bounded \(0 < I_s < \lim_{N \to \infty} H_{N,s} < 1 + I_s \), with a simple integral \(I_s = \int_1^{\infty} x^{-s} {\mathrm{d}}x\). 
The integral diverges logarithmically for \(s=1\) whereas \(\int_1^{N} x^{-s} {\mathrm{d}}x = N^{1-s}/(1-s) - 1/(1-s)\) for \(0<s<1\). 
The Anderson localization criterion then gives \(\lim_{N\to\infty} S_1(N) = \sqrt{2/\pi} \bigl[ 1 + 1/\sqrt{2} \, \zeta(s) \bigr]\) for \(s > 1\), whereas the ergodic criterion is satisfied for \(s \le 1/2\). 
The eigenvector distribution in the region \(1/2 < s < 1\) was shown~\cite{bogomolny2018power} to have no anomalous scaling, implying that the fractal dimensions are the same as for the non-banded random matrices and thus that this phase is ergodic.

\bibliography{mbl, general, glass, software, ml, local}

\end{document}